\documentclass[aps,twocolumn,pre,floatfix,superscriptaddress]{revtex4-2}
\usepackage{amssymb}
\usepackage{amsmath}
\usepackage{bm}
\usepackage{graphicx}
\usepackage{dcolumn}
\usepackage{longtable}
\usepackage{color} 
\usepackage{xcolor}
\definecolor{gray}{RGB}{110,110,110}
\usepackage[mathscr]{euscript}
\usepackage{xr}

\usepackage{titlesec}
\titleformat{\section}{\normalfont\sffamily\bfseries}{\thesection}{0pt}{}
\titlespacing{\section}{0pt}{12pt}{0pt}
\titleformat{\subsection}[runin]{\normalfont\bf}{\thesection}{0pt}{}[. ]

\renewcommand{\figurename}{\textsf{\textbf{{Figure}}}}

\newcommand{\bmm}[1]{{\bf #1}}  

\newcommand{\swap}[2]{#1}

\makeatletter
\def\maketitle{
\@author@finish
\title@column\titleblock@produce
\suppressfloats[t]}
\makeatother

\newcolumntype{C}[1]{>{\hfil}m{#1}<{\hfil}}

\begin{document}

\title{Compass-model physics on the hyperhoneycomb lattice \\in the extreme spin-orbit regime}

\author{Ryutaro~Okuma}
\email{ryutaro.okuma@physics.ox.ac.uk}
\affiliation{Clarendon Laboratory, University of Oxford Physics Department, Parks Road, Oxford OX1 3PU, UK
}%
\affiliation{%
 Institute for Solid State Physics, University of Tokyo, Kashiwa, Chiba 277-8581, Japan
}%
\author{Kylie~MacFarquharson}
\affiliation{Clarendon Laboratory, University of Oxford Physics Department, Parks Road, Oxford OX1 3PU, UK
}%
\author{Roger~D.~Johnson}
\affiliation{Department of Physics and Astronomy, University College London, Gower Street, London WC1E 6BT, UK}
\author{David~Voneshen}
\affiliation{ISIS Facility, Rutherford Appleton Laboratory, Chilton, Didcot OX11 0QX, UK}
\affiliation{Department of Physics, Royal Holloway University of London, Egham, TW20 0EX, UK}
\author{Pascal~Manuel}
\affiliation{ISIS Facility, Rutherford Appleton Laboratory, Chilton, Didcot OX11 0QX, UK}
\author{Radu~Coldea}
\affiliation{Clarendon Laboratory, University of Oxford Physics Department, Parks Road, Oxford OX1 3PU, UK
}%

\date{\today}
\maketitle

\section*{Abstract}
The physics of spin-orbit entangled magnetic moments of $4d$ and $5d$ transition metal ions on a honeycomb lattice has been much explored in search for unconventional magnetic orders or quantum spin liquids expected for compass spin models, where different bonds in the lattice favour different orientations for the magnetic moments. Realizing such physics with rare-earth ions is a promising route to achieve exotic ground states in the extreme spin orbit limit, however this regime has remained experimentally largely unexplored due to major challenges in materials synthesis. Here we report successful synthesis of powders and single crystals of $\beta$-Na$_2$PrO$_3$, with $4f^{1}$ Pr$^{4+}$ $j_\mathrm{eff}\!=\!1/2$ magnetic moments arranged on a hyperhoneycomb lattice with the same threefold coordination as the planar honeycomb. We find a strongly noncollinear magnetic order with highly dispersive gapped excitations that we argue arise from frustration between bond-dependent, anisotropic off-diagonal exchanges, a compass quantum spin model not explored experimentally so far. Our results show that rare-earth ions on threefold coordinated lattices offer a platform for the exploration of quantum compass spin models in the extreme spin orbit regime, with qualitatively distinct physics from that of $4d$ and $5d$ Kitaev materials.

\section*{Introduction}
\begin{figure}
\includegraphics[width=8cm]{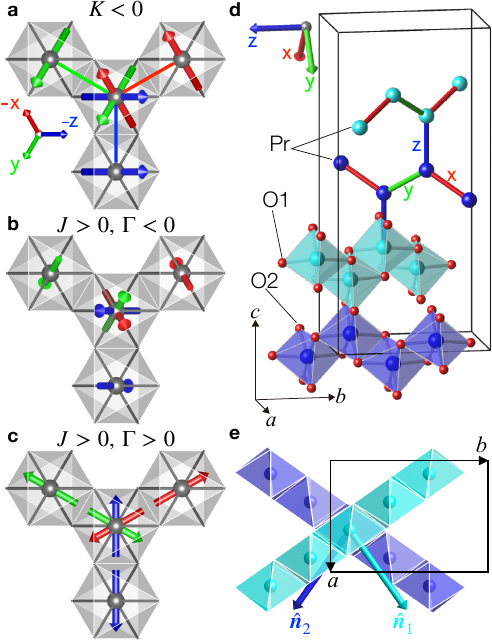}
\caption{\label{fig:structure} \textbf{Frustration from different bond-dependent anisotropic exchanges on the hyperhoneycomb lattice.} \textbf{a,} Local threefold coordination of Pr$^{4+}$ ions (grey balls) inside edge-sharing cubic O$_6$ octahedra (gray shading). Spin pairs (arrows) for each bond are colour coded red/green/blue according to the label $\sf{xyz}$ of the axis normal to the Pr-O$_2$-Pr superexchange plane of that bond. In the Kitaev model with $K<0$ each bond prefers the spins at the two ends to point along the direction normal to the superexchange plane, so the central spin is frustrated among the three orthogonal $\sf{xyz}$ directions. \textbf{bc,} The $\Gamma$ interaction also creates frustration, but among other spin directions. \textbf{b,} $J\Gamma$ model for Heisenberge exchange $J>0$ and $\Gamma<0$. Each bond prefers spins at the two ends antiparallel, in the superexchange plane and orthogonal to the bond direction. The central spin is frustrated among three non-coplanar directions at 60$^{\circ}$ relative to each other. \textbf{c,} Same as \textbf{b,} but for $\Gamma>0$, where each bond prefers spins at the two ends to be antiparallel and along the bond direction. The central spin is frustrated among three coplanar directions at 60$^{\circ}$ relative to each other. \textbf{d,} Crystal structure of $\beta$-Na$_2$PrO$_3$ (orthorhombic $Fddd$ space group, see Supplementary Note 2 \cite{SM}) showing the threefold coordinated hyperhoneycomb lattice of Pr$^{4+}$ ions (dark/light blue balls) located inside edge-sharing O$_6$ octahedra forming zigzag chains that alternate in direction between the two basal plane diagonals $\bmm{a}\pm\bmm{b}$ (Na ions not shown for brevity). Bonds are colour coded and labelled as in \textbf{a} for an ideal structure with cubic octahedra. \textbf{e,} Light/dark octahedral zigzag chains in d) projected onto the basal $ab$ plane. Coloured arrows labelled $\bmm{\hat{n}}_{1,2}$ show normals to the distinct planes defined by the light/dark zigzag chains. This noncoplanarity enhances the frustration effects of $\Gamma$ interactions.}
\end{figure}

Materials with magnetic moments interacting via bond-dependent anisotropic interactions are attracting much attention as candidates to display novel cooperative behaviour. This is exemplified by the Kitaev model on the honeycomb lattice for spin-1/2 moments illustrated in Fig.~\ref{fig:structure}a, where Ising exchanges $KS_i^\gamma S_j^\gamma$ with $\gamma={\sf x, y, z}$ couple mutually orthogonal spin components for the three bonds sharing a common site, with the resulting strong frustration stabilising an exactly solvable quantum spin liquid \cite{kitaev2006anyons}. Candidates to realize such physics are heavy transition metal ions, such as $4d^5$ Ru$^{3+}$ or $5d^5$ Ir$^{4+}$ ions, located inside edge-sharing cubic octahedra, where the combination of spin-orbit coupling and crystal field stabilize magnetic moments with mixed spin-orbital character, that can interact via bond-dependent anisotropic exchanges of predominant Kitaev character \cite{chaloupka2010kitaev}. Experimental studies of candidate materials \cite{takagi2019concept} have revealed novel phenomena such as spin-momentum locking in Na$_2$IrO$_3$ \cite{hwan2015direct}, unconventional continuum of excitations \cite{banerjee2017neutron} and thermal transport \cite{kasahara2018majorana,yokoi2021half} in $\alpha$-RuCl$_3$, and counterrotating incommensurate orders in $\alpha$-, $\beta$- and $\gamma$-Li$_2$IrO$_3$ \cite{williams2016,biffin2014unconventional,biffin2014noncoplanar}.

An equally important yet distinct bond-dependent anisotropic interaction is the off-diagonal symmetric exchange $\Gamma$ \cite{rau2014generic}, which couples spin components normal to the Kitaev axes, i.e. $\Gamma(S_i^{\sf x}S_j^{\sf y}+S_i^{\sf y}S_j^{\sf x})$ for a ${\sf z}$-bond. Such terms also generate frustration as each spin is conflicted into pointing along incompatible directions favoured by the three bonds sharing that site, see Fig.~\ref{fig:structure}b, with the set of directions changing upon reversing the sign of $\Gamma$ (see Fig.~\ref{fig:structure}c), with both sets of directions different from those favoured by a Kitaev term as illustrated in Fig.~\ref{fig:structure}a. The $\Gamma$ model on the honeycomb lattice has a macroscopically degenerate manifold of classical ground states \cite{PhysRevLett.118.147204} and the quantum model is not exactly solvable. Symmetry-protected topological phases have been predicted in a honeycomb ladder with $\Gamma$ and Heisenberg exchange \cite{JG_ladder}.

Here we report experimental studies that find evidence for substantial $\Gamma$ interactions in $\beta$-Na$_2$PrO$_3$ \cite{wolf19882} where Pr$^{4+}$ $4f^{1}$ ions form a three-dimensional hyper-honeycomb lattice with the same local threefold coordination as the planar honeycomb. Pr$^{4+}$ ions have long been theoretically predicted to realize quantum compass spin models with potentially different Hamiltonians compared to the heavy transtion metal ions Ru$^{3+}$ and Ir$^{4+}$ due to the stronger spin orbit coupling and the different characteristics of the orbitals involved in superexchange \cite{motome2020materials,jang2020computational}. However, up to now no physical properties have been reported for $\beta$-Na$_2$PrO$_3$ because the synthesis is hampered by severe air-sensitivity and the presence of more stable polymorphs \cite{hinatsu2006crystal,daum2021collective}.

The three-dimensional hyperhoneycomb lattice is an ideal playground to explore frustration effects from $\Gamma$ interactions. It has an orthorhombic unit cell with zigzag chains running alternatingly along the $\bmm{a}\pm\bmm{b}$ basal plane directions with vertical bonds connecting the two types of chains (see Fig.~\ref{fig:structure}d) such that each site has thee nearest neighbours. This magnetic lattice has been realized experimentally so far only in $\beta$-Li$_2$IrO$_3$ \cite{takayama2015hyperhoneycomb} and $\beta$-ZnIrO$_3$ \cite{haraguchi2022quantum}, in the latter case with chemical disorder on the Zn site. In the absence of a $\Gamma$ term, there is no critical difference between the planar honeycomb and the hyperhoneycomb: the Kitaev model on both lattices has exactly solvable quantum spin liquid ground states \cite{PhysRevB.79.024426}, and four types of similar collinear magnetic structures appear at the same value of $K/J$ at the mean field level \cite{rau2014generic,lee2014heisenberg,lee2015hyperhoneycomb}, where $J$ is the Heisenberg exchange. Introduction of $\Gamma$ for the hyperhoneycomb lattice renders most of the phases noncollinear and even noncoplanar \cite{lee2015hyperhoneycomb}, whereas noncollinear orders are realized in the planar honeycomb only when $K$ and $\Gamma$ are dominant \cite{rau2014generic}. The key difference is attributed to the fact that in the hyperhoneycomb structure the zigzag chains are contained within distinct planes as illustrated in Fig.~\ref{fig:structure}e. This intrinsic non-coplanarity enhances the frustration effects from $\Gamma$ interactions as one cannot define a global, common plane for all the spins, unlike the case of the two-dimensional planar honeycomb where all bonds are coplanar. Furthermore, in the coplanar case a trigonal compression along the direction ($Z$) normal to the honeycomb layer can lead to bond-independent XXZ-type interactions, as in the case of the Co-based honeycomb BaCo$_2$(AsO$_4$)$_2$ \cite{halloran2023geometrical}. An Ising-like XXZ model has also been proposed to describe the layered honeycomb $\alpha$-Na$_2$PrO$_3$ \cite{daum2021collective}. In contrast, such a model is not  applicable to the hyperhoneycomb lattice due to its twisted 3D connectivity, with any anisotropy originating instead from bond-dependent anisotropic exchanges.

\begin{figure}
\includegraphics[width=8.5cm]{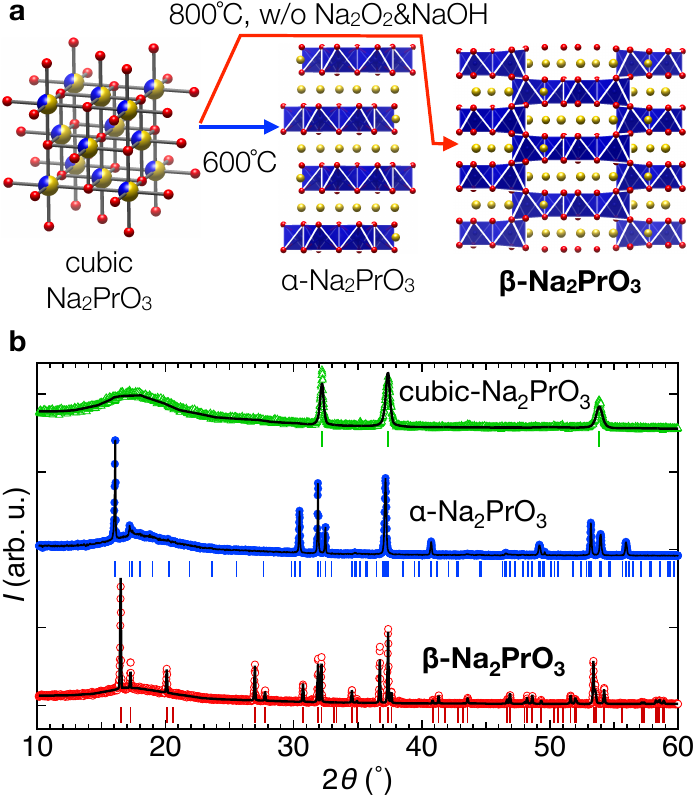}
\caption{\label{fig:synthesis}\textbf{Synthesis of hyperhoneycomb $\beta$-Na$_2$PrO$_3$.} \textbf{a,} Phase pure cubic-Na$_2$PrO$_3$ synthesized at 400$^\circ$C is further annealed at 800$^\circ$C to obtain phase pure $\beta$-Na$_2$PrO$_3$. Annealing at a lower temperature of 600$^\circ$C stabilises $\alpha$-Na$_2$PrO$_3$. Red, blue, yellow spheres indicate oxygen, praseodymium, and sodium atoms, respectively. \textbf{b,} XRD powder diffraction patterns (monochromatic Cu-K$_{\alpha1}$) of the three polymorphs of Na$_2$PrO$_3$, offset vertically for clarity. Red, blue and green symbols show the diffraction pattern of $\beta$-, $\alpha$-, and cubic-Na$_2$PrO$_3$, respectively. Black lines and bars under each data set indicate Rietveld refinement fits and positions of Bragg peaks for each phase, $\alpha$- from Ref.~\cite{powderalphaNa2PrO3}, $\beta$- from Supplementary Table~\swap{\ref{tab:XRD}}{I}, and cubic-phase from Ref.~\cite{wolf19882}.}
\end{figure}

{\em Findings of this study} We have succeeded in selective synthesis of phase-pure powders and sizeable single crystals of $\beta$-Na$_2$PrO$_3$, which realises a hyperhoneycomb lattice with $j_{\rm eff}=1/2$ Pr$^{4+}$ magnetic moments. A critical insight that enabled the synthesis was understanding the role played by the melting species Na$_2$O$_2$ and NaOH in the chemical stability of the various polymorphs of Na$_2$PrO$_3$, see Fig.~\ref{fig:synthesis}. By combining neutron diffraction and inelastic scattering with magnetic symmetry analysis and spinwave calculations we obtain a full solution of a highly noncollinear magnetic structure with gapped and strongly dispersive spin-wave excitations. We provide evidence that this physics is governed by frustrated bond-dependent anisotropic interactions, but of a different character from the much explored Kitaev exchange.

\section*{Results}
\begin{figure*}
\includegraphics[width=175mm]{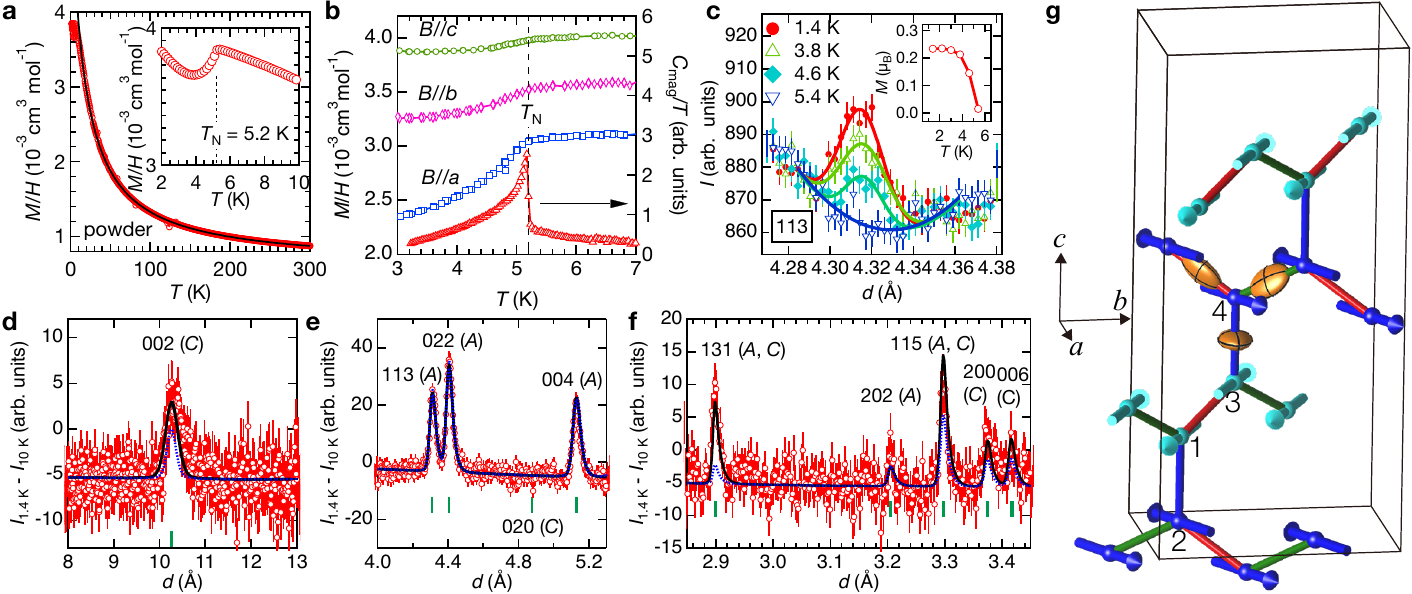}
\caption{\label{fig:diffraction}\textbf{Noncollinear magnetic order in $\beta$-Na$_2$PrO$_3$}. \textbf{a,} Temperature dependence of the magnetic susceptibility (red circles) under field-cooling (FC) of 1~T. No difference is observed between FC and zero-FC. The black line is a fit to a Curie-Weiss form in the region 20 to 300~K. (Inset) Susceptibility anomaly near the onset of magnetic ordering. \textbf{b,} Temperature dependence of single-crystal susceptibility and ac heat capacity of a multi-crystal sample. Blue squares, pink diamonds, green circles and red triangles indicate the susceptibility in field applied along the $a$, $b$ and $c$ axes, and zero-field heat capacity (right-hand axis), respectively. The heat capacity points are raw data with an estimate of the addenda contribution subtracted off.  \textbf{c,} Temperature dependence of the raw neutron diffraction intensities near the magnetic Bragg peak position (113) with an almost absent nuclear contribution, solid lines are fits to Gaussian peaks with a second order polynomial as a background determined from a fit to the data at 5.2 K. Inset shows the temperature dependence of the fitted magnitude of the ordered magnetic moment (solid line is guide to the eye).\textbf{d,} Magnetic Bragg peak at the nuclear forbidden position (002) observed in the bank that covers the largest $d$-spacing. Data points are the intensity difference between base temperature (1.4~K) and paramagnetic (10~K), and the green vertical bar under the pattern shows the nominal magnetic Bragg peak position. Solid black/dotted blue lines are fits to the ($A_x,\mp C_y$) models described in the text. Symbols and lines have the same meaning in panels \textbf{e} and \textbf{f}, which show magnetic reflections with lower $d$-spacing. Error bars on data points in panels c-f represent one standard deviation. \textbf{g,} Schematic diagram of the strongly noncollinear magnetic structure ($A_x,-C_y$) refined from neutron powder diffraction data (for projections onto different crystallographic planes see Supplementary Fig.~\swap{\ref{fig:magnetic_structure}}{6}). Labels 1-4 indicate sites equivalent to those in the primitive cell (listed in Supplementary Table~\swap{\ref{tab:BV}}{VI}) up to $F$-centring translations. The anisotropy of the bond dependent interactions is illustrated by the yellow ellipsoids with principal planes indicated by black contours. For the zigzag bonds the ellipsoids are elongated along the bonds, on the vertical bonds they are elongated transverse to the bonds, along the $a$-axis with the largest antiferromagnetic ordered components. For better visualisation of the ellipsoid's shape, we used $\Gamma/J$ and $\Gamma'/J$ twice as large as the estimated values from fitting the magnetic excitation spectrum.}
\end{figure*}

{\em Magnetic susceptibility.} We first characterize the magnetic behaviour using powder magnetic susceptibility measurements plotted in Fig.~\ref{fig:diffraction}a. The data can be fitted in the region of 20 to 300~K by a Curie Weiss law $\chi(T) = \chi_0 + \mu_0\mu_\mathrm{eff}^2/3k_\mathrm{B}(T - T_\mathrm{CW})$ with $\chi_0 \!=\! 6.12(9)\times 10^{-4}\mathrm{cm}^3\ \mathrm{mol}^{-1}$, $T_\mathrm{CW} \!=\! -15(1)$ K indicating overall antiferromagnetic interactions, and $\mu_\mathrm{eff} \!=\! 0.81(1)\mu_\mathrm{B}$/Pr, which implies a $g$-factor $g \!=\! 0.94(1)$, smaller than the $10/7$ value expected in the limit of very weak cubic crystal field \cite{harris}. Such moment reduction was also observed for $\alpha$-Na$_2$PrO$_3$ \cite{daum2021collective} and attributed to mixing with higher crystal field levels. Fig.~\ref{fig:diffraction}a(inset) shows a clear anomaly at $T_\text{N}\!=\!5.2$~K, attributed to the onset of magnetic order. This is more clearly seen in susceptibility measurements on single crystals in Fig.~\ref{fig:diffraction}b, where a sudden decrease in the susceptibility along the orthorhombic $a$-axis is observed below the same temperature as in the powder data, as expected for the onset of a magnetic structure with dominant antiferromagnetic $a$-axis components. Heat capacity (Fig.~\ref{fig:diffraction}b lower trace) further corroborates this scenario by observing a very sharp peak at the same temperature as the kink in susceptibility.

{\em Magnetic propagation vector.} To determine the magnetic structure we use neutron powder diffraction, which revealed new diffraction peaks as well as an intensity increase at some structural peak positions upon cooling below $T_\text{N}$. The temperature dependence of the diffraction intensity is illustrated in Fig.~\ref{fig:diffraction}c, which shows the intensity at a position where the structural Bragg diffraction is almost absent (as explained in Supplementary Note 3). At base temperature a clear peak is observed, which decreases monotonically upon heating and can no longer be detected at $T_\text{N}$. The magnetic peaks are most clearly revealed in the difference pattern between base temperature (1.4~K) and paramagnetic (10~K), shown in panels d,e,f, where green bars under the pattern indicate the nominal peak positions for the F-centered lattice. Fig.~\ref{fig:diffraction}d reveals a structurally-forbidden peak at (002) around a $d$-spacing of 10~\AA, which breaks the selection rule for $(00l)$ structural reflections with $l \!=\! 4p$ ($p$ integer) characteristic of the structural $Fddd$ space group (due to the $d$ diamond glides normal to the $a$ and $b$ axes). Over 10 magnetic Bragg peaks could be detected in total and all could be indexed by all-odd or all-even Miller indices, indicating that the magnetic and structural unit cells are the same, i.e. the magnetic propagation vector is $\bmm{q}\!=\!\bf{0}$.

{\em Magnetic basis vectors.} Symmetry analysis showed that any given magnetic structure can be decomposed into a linear combination of 12 modes: $F_i$, $A_i$, $C_i$, and $G_i$, where $i$ denotes the polarisation of the mode (along $xyz$ axes, defined to be along the orthorhombic $abc$ axes) and $F$, $A$, $C$ and $G$ denote basis vectors, which encapsulate symmetry-imposed relations between the moment orientations (parallel or antiparallel) at the four Pr sites in the primitive cell as listed in Supplementary Table~\swap{\ref{tab:BV}}{VI}. $F$ means ferromagnetic, $A$ nearest-neighbour antiferromagnetic, $C$ has parallel (antiparallel) spins on the vertical (zigzag) bonds, and $G$ viceversa. Each basis vector satisfies distinct selection rules (summarized in Supplementary Table~\swap{\ref{tab:reflection}}{VII}) with the consequence that simply the presence of certain magnetic Bragg peaks uniquely identifies the presence of certain basis vectors. In addition, the fact that magnetic neutron scattering is only sensitive to the magnetic moment components perpendicular to the wavevector transfer, provides further information to identify the polarization of the basis vectors. Finite intensity for the pure $C$-peaks (006) and (200) in Fig.~\ref{fig:diffraction}f and absence of a peak at the pure $C$ position (020) in Fig.~\ref{fig:diffraction}e means that a basis vector $C$ must be present and it must be polarized along the $y$-axis. Finite intensity at the mixed $AF$ position (004) in Fig.~\ref{fig:diffraction}e identifies the second basis vector as $A_x$, since $A_y$ would lead to the unphysical situation of unequal moment magnitudes on the four sites in the primitive cell, and any $F$ (ferromagnetic) component can be ruled out by the absence of a ferromagnetic anomaly at $T_\text{N}$ in the magnetization data in Fig.~\ref{fig:diffraction}a(inset). In the following we show that the basis vectors $A_x$ and $C_y$ found by inspection of the neutron powder diffraction pattern can describe it fully quantitatively, resulting in a complete solution of the magnetic structure.

{\em Full magnetic structure.}
A magnetic structure with only $A_x$ and $C_y$ basis vectors corresponds to a single irreducible representation (irrep), $m\Gamma_4^-$, for symmetry-allowed $\bmm{q}\!=\!\bf{0}$ magnetic structures (tabulated in Supplementary Table~\swap{\ref{tab:MSG}}{VIII}), consistent with a continuous transition at $T_\text{N}$, as observed in the heat capacity data in Fig.~\ref{fig:diffraction}b. The magnitudes $M_x$ and $M_y$ of the two basis vectors can be separately determined from the intensity of magnetic Bragg peaks where only one of them contributes, such as (004) for $A_x$ and (002) for $C_y$. The relative phase between the two basis vectors can be determined from the intensity of magnetic Bragg peaks where both basis vectors contribute, as the total intensity is the sum of the intensities due to the two separate basis vectors, plus an additional cross-term that is sensitive to the relative phase (for details see Supplementary Note 2). The magnetic diffraction pattern contains many magnetic Bragg peaks that can be used for this purpose, all peaks with $hkl$ all-odd have contributions from both $A$ and $C$ basis vectors. Symmetry analysis predicts that the two basis vectors can be in-phase ($A_x,C_y$), or in antiphase ($A_x,-C_y$), and these two scenarios can be differentiated by neutron diffraction, as we will show later. The in-phase/antiphase structures have ordered spins oriented predominantly perpendicular (parallel) to the direction of the zigzag chains with the exact orientation being a degree of freedom related to the relative magnitudes of $M_x$ and $M_y$.

We have tested both scenarios, freely refining $M_x$ and $M_y$ in the fits. The best fit to an antiphase ($A_x,-C_y$) structure gives a very good account of the observed pattern, all peaks in Fig.~\ref{fig:diffraction}d-f panels are quantitatively accounted for (black solid line, for more details of the fits see Supplementary Note 2). In contrast, the alternative fit (dashed blue) to an in-phase ($A_x,C_y$) structure cannot fully account for the intensity at well measured peaks such as (002) in Fig.~\ref{fig:diffraction}d and (131), (115) in panel f. In the best fit structure the total ordered moment at each site is $M\!=\!\sqrt{M_x^2+M_y^2}\!=\!0.222(4)\mu_\text{B}$ with the relative ratio $M_y/M_x\!=\!0.55(2)$. The resulting magnetic structure has spins confined to the $ab$ plane, close (at an angle of $26(1)^{\circ}$) to the direction of the zigzag chains that they belong to. The resulting global magnetic structure is strongly noncollinear, as moments belonging to zigzag chains connected by a vertical bond make a relatively large angle $180^\circ-2\arctan(M_y/M_x)\!=\!123(2)^\circ$, see Fig.~\ref{fig:diffraction}g. The refined magnetic structure naturally explains the dominant features in the temperature-dependent susceptibility data in Fig.~\ref{fig:diffraction}b, which observed a prominent suppression below $T_\text{N}$ of the susceptibility along the $a$-axis, the direction with the dominant antiferromagnetic components, contrasting with an almost constant susceptibility below $T_\text{N}$ for field along the $c$-axis, normal to the plane of the ordered spins.

\begin{figure*}
\includegraphics[width=175mm]{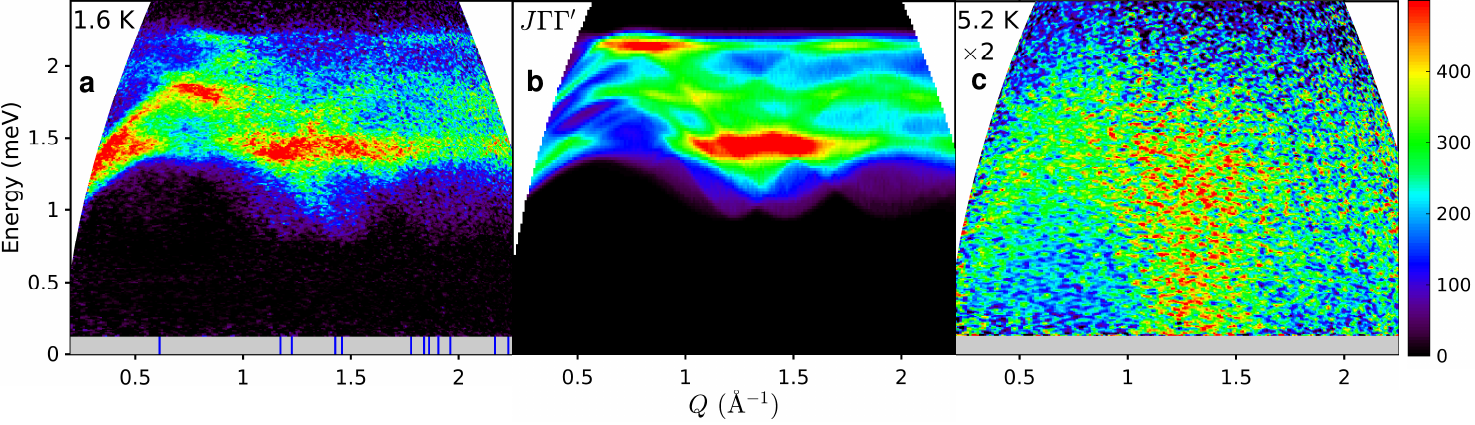}
\caption{\label{fig:spinwave} \textbf{Powder inelastic neutron scattering spectrum.} \textbf{a,} Base temperature data (1.6~K, $E_i\!=\!3.7$~meV) observing dispersive magnetic excitations above a spin gap. Thin vertical blue bars at the bottom of the panel indicate wavevectors of magnetic Bragg peaks of the determined magnetic structure. Gray shading shows the inaccessible region close to the elastic line dominated by incoherent elastic scattering. \textbf{b,} Spherically-averaged spinwave spectrum of the minimal $J\Gamma\Gamma'$ model described in the text, convolved with the estimated experimental energy resolution. \textbf{c,} Same as a, but at 5.2~K and with intensities scaled by $\times2$. The enhanced inelastic signal in a broad momentum range near 1.3~\AA$^{-1}$ extending down to the lowest energies is attributed to precursor dynamical correlations from which the magnetic order develops upon cooling. The colour bar indicates scattering intensity in arbitrary units on a linear scale.}
\end{figure*}

{\em Inelastic Neutron Scattering.}
To gain insight into the interactions that could stabilize the determined noncollinear structure we performed powder inelastic neutron scattering measurements. The observed magnetic excitation spectrum deep in the ordered state is shown in Fig.~\ref{fig:spinwave}a as a function of energy and wavevector transfer $Q$. The spectrum has a clear gap of around 0.75~meV and extends up to 2.3~meV, being highly structured with a combination of highly-dispersive features and prominent near-flat regions. As expected, the energy scale of the spectrum is comparable with the Curie-Weiss temperature of $-k_\mathrm{B}T_\mathrm{CW}\simeq1.3$~meV. The lower boundary of the spectrum is highly dispersive with clear gapped minima seen for $Q$ near 1.25 and 1.45~\AA$^{-1}$, which are close to wavevector positions where several magnetic Bragg peaks occur in the diffraction pattern, indicated by thin vertical blue bars at the bottom of the panel. The magnetic character of the observed spectrum is further confirmed by measurements at $T_\mathrm{N}$ shown in Fig.~\ref{fig:spinwave}c, the gaps fill in and the dispersive and strong intensity features observed at lower temperature wash out.

{\em Spin gap.} The substantial spin gap indicates a considerable energy cost in moving the spins away from their local orientations in the magnetic ground state. The presence of such energetically strongly-preferred directions cannot be due to single-ion anisotropy effects. For a Kramers ground state doublet as expected for $4f^1$ Pr$^{4+}$ ions in octahedral crystal field \cite{jang2020computational}, there cannot be any local, single-ion anisotropy terms, as all even powers of the components of the effective angular momentum of 1/2 describing the ground state doublet are constants \cite{Yoshida}. The substantial gap observed therefore must be due to anisotropic exchange interactions, as we discuss below.

{\em Spin Hamiltonian.}
The non-collinear magnetic order observed could in principle be stabilized by the competition between just two exchange terms, a nearest-neighbour antiferromagnetic Heisenberg exchange $J$ and a symmetry-allowed Dzyaloshinskii-Moriya interaction $D$ on the vertical bonds. However, to quantitatively explain the large angle between moments in the $ab$ plane would require an unphysically large $D/J$ ratio and also such a model would have a gapless spectrum, in contrast with the substantial gap seen experimentally (for more details see Supplementary Note~5). We therefore consider the origin of the observed magnetic structure and spin wave excitations within a $JK\Gamma$ model, expected microscopically via a strong-coupling expansion of a multiband tight-binding model with on-site interaction and large atomic spin orbit coupling, in this framework $K$ and $\Gamma$ appear in the presence of a Hund's coupling \cite{jang2020computational,lee2015hyperhoneycomb}. Such a model with $J>0$, $K\!=\!0$ and $\Gamma<0$ can qualitatively reproduce several of the key features observed experimentally. First, it gives a noncollinear magnetic ground state with moments confined in the $ab$ plane, with the same basis vectors $A_x$ and $C_y$ found experimentally, and with $A_x$ dominant. However, the structure is predicted to be the in-phase combination ($A_x,C_y$) (labelled AF$_a$ in \cite{lee2015hyperhoneycomb}), not the antiphase ($A_x,-C_y$) found experimentally. Second, it naturally gives a spin gap, scaling as $\sqrt{-\Gamma J}$ at leading order in $\Gamma/J$.

The antiphase basis vector combination ($A_x,-C_y$) with $A_x$ dominant is not contained in the phase diagram of the above $JK\Gamma$ model \cite{lee2015hyperhoneycomb}, however it can be stabilised by relaxing the assumption that all bonds are symmetry-related. Indeed, in the $Fddd$ crystal structure the zigzag bonds are not symmetry-equivalent to the vertical bonds. By calling the off-diagonal exchange on the zigzag bonds $\Gamma'$ and making $\Gamma^{\prime}>0$ on the $\sf{x}$-bond of the chains running along $\bmm{a}-\bmm{b}$ (such as 1-3 in Fig.~\ref{fig:diffraction}g) stabilises the basis vector combination ($A_x,-C_y$) as found experimentally. This suggests that the symmetry-inequivalence of the zigzag and vertical bonds is the most likely reason for the observed antiphase basis vector combination in the ground state.

To illustrate the level of agreement that can be obtained by such a minimal model, we assume for simplicity equal magnitude and opposite sign off-diagonal exchanges on the zigzag and vertical bonds, i.e. $\Gamma'\!=\!-\Gamma$ and perform a fit of the observed powder-averaged spectrum freely varying $J$ and $\Gamma$, this gives $J\!=\!1.22(4)$~meV and $\Gamma\!=\!-0.27(5)$~meV. This minimal model reproduces reasonably quantitatively the dominant highly-dispersive features and regions of strong intensity in the powder-averaged spectrum, compare Fig.~\ref{fig:spinwave}a with b (for further details of the mean-field analysis and spinwave calculations see Supplementary Notes 4-6). We take this agreement as indication that the bond-dependent $\Gamma$ and $\Gamma'$ interactions are the most relevant subleading exchanges after the isotropic Heisenberg term $J$. In the above minimal model, the mean-field calculated $M_y/M_x\!=\!0.14$ (assuming an isotropic $g$-tensor in the $ab$ plane) is smaller than the ratio 0.55 deduced experimentally, also there are differences of detail between the observed and calculated powder spectrum in Fig.~\ref{fig:spinwave}a and b, we attribute those differences to extensions of the Hamiltonian beyond the minimal model considered here, which however we do not expect would change the physics qualitatively, but could improve the level of quantitative agreement with the experiment.

{\em Ground state selection.} In the minimal model considered above the ground state selection occurs as follows. An antiferromagnetic Heisenberg exchange $J>0$ on all bonds selects collinear N\'{e}el order, adding off-diagonal exchange $\Gamma<0$ on the vertical bonds breaks the spherical rotational symmetry, opens a gap in the spectrum and selects the $a$-axis for the moments' direction in the ground state (basis vector $A_x$). Adding now off-diagonal exchange $\Gamma'>0$ on the zigzag bonds rotates the moments at the two ends of a vertical bond in opposite senses, but keeps all spins in the same zigzag chain collinear and antiferromagnetically ordered by mixing in a $-C_y$ basis vector in the ground state; this is such as to bring the spins towards locally-favoured easy directions in each chain, with those directions different between chains connected by a vertical bond and related by a $2_z$ rotation. Therefore, the non-collinearity of the magnetic order can be understood as a direct consequence of the frustration between the off-diagonal exchanges on the different bonds.

\section*{Discussion}
The magnetic behaviour of $\beta$-Na$_2$PrO$_3$ is quite different from that of isostructural $\beta$-Li$_2$IrO$_3$ \cite{takayama2015hyperhoneycomb}, the only other known material with a magnetic hyperhoneycomb lattice with no structural disorder. The latter material has a non-coplanar, incommensurate magnetic structure with counter-rotating moments \cite{biffin2014unconventional}, in contrast $\beta$-Na$_2$PrO$_3$ has a non-collinear, commensurate magnetic structure. The underlying spin Hamiltonians are qualitatively different, in $\beta$-Li$_2$IrO$_3$ a dominant $K$ has been proposed \cite{halloran2022}, whereas in $\beta$-Na$_2$PrO$_3$ we find a $J\Gamma\Gamma'$ model, we attribute this difference to the distinct orbitals and superexchange mechanisms involved in the two cases. The presence of a clear spin gap in $\beta$-Na$_2$PrO$_3$ of magnitude comparable to the Zeeman energy of accessible applied magnetic fields opens up the prospect of observing experimentally novel field-induced magnetic phases and unconventional spin dynamics, such as topological nodal lines and Weyl magnons, protected by magnetic glide symmetries and arising from magnon pairing effects, theoretically predicted \cite{choi2019nonsymmorphic} for generic bond-dependent anisotropic Hamiltonians on the hyperhoneycomb lattice, but not yet observed experimentally.

By incorporating rare-earth $4f$ ions in threefold coordinated lattices, with orbitals of different character mediating the superexchange interactions in the extreme spin-orbit regime, we have been able to access a different materials platform for realizing quantum compass spin models with largely distinct physics from the much explored 4$d$ and 5$d$ Kitaev materials. We have facilitated the first steps by resolving materials synthesis challenges and observed new physics driven by frustrated off-diagonal exchanges.

Beyond $\beta$-,$\gamma$-Li$_2$IrO$_3$ and $\beta$-Na$_2$PrO$_3$, realization of three-dimensional threefold coordinated lattices in more systems will be an important task to reveal the rich physics of the quantum compass models. Notable progress in this direction is the recently demonstrated control of cation ordering in the rock salt structure, used to obtain a three-dimensional network of corner- and edge-sharing octahedra in Li$_3$Co$_2$SbO$_6$ \cite{brown2019synthesis, duan2022anomalous}. The use of high pressure could also potentially transform layered honeycomb materials into 3D hyperhoneycomb lattices, as illustrated by high-pressure studies on IrI$_3$\cite{ni2022honeycomb}. Very recently, an organic molecule was employed to realize another threefold coordinated lattice of much theoretical interest, the hyperoctagon lattice, in a Co-based metal organic framework \cite{ishikawa2024j}. We anticipate that these and other novel synthetic procedures will expand the materials platform of three-dimensional threefold coordinated lattices and allow a wider experimental exploration of the rich range of cooperative magnetic behaviours expected for such geometries in the presence of strong spin orbit coupling, as we have reveled in hyperhoneycomb $\beta$-Na$_2$PrO$_3$.

\section*{Methods}
{\em Synthesis.} Full details of the synthesis using a solid-state reaction protocol under inert atmosphere are provided in Supplementary Note 1.

{\em Crystal structure determination.} The $\beta$-phase crystal structure was determined via single-crystal x-ray diffraction using a Mo source SuperNova diffractometer on a crystal with dimensions $170 \times 95 \times 31$~$\mu\text{m}^3$, covered with vacuum grease to protect it from air. No evidence for sample degradation was observed within the duration of the x-ray measurements (less than a couple of hours).

{\em Magnetic characterization.} Magnetization measurements were performed using a Quantum Design MPMS3 system in fields up to 7~T and temperatures down to 2~K, first on powders of typical mass 22.7~mg and subsequently on co-aligned single crystals with a total combined mass of order 0.2~mg. The single crystals were initially handled in vacuum pump oil, which was removed by washing with toluene. For the magnetization measurements in field along specific crystallographic directions ($a$,$b$,$c$) the crystals were aligned and fixed onto a flat plate (single crystal of NaCl with typical dimensions $1.4 \times 1.3 \times 0.7$ mm$^3$). Melted paraffin wax was used to coat the crystals and fix them onto the flat plate and an aluminium foil was attached below the flat plate to almost cancel the diamagnetic contribution of paraffin wax. After the measurements along all the axes, the crystals were removed by melting the wax and the background susceptibility signal, comprising of diamagnetism from wax and NaCl, and paramagnetism of the aluminium foil, was measured and subtracted off to obtain the intrinsic Na$_2$PrO$_3$ susceptibility. Both the powder and single crystal susceptibility measurements showed evidence for a small ferromagnetic impurity, identified as PrO$_{2-x}$ due to decomposition. The contribution of this impurity was subtracted by comparing the magnetisation data to single-crystal torque data (to be described in detail elsewhere).

{\em Heat capacity.} Zero-field heat capacity was performed on a collection of single crystals of combined mass $0.12(2)$~mg using a custom ac heat capacity setup operating with a Quantum Design PPMS system.

{\em Neutron diffraction.} Neutron powder diffraction measurements were performed using the time-of-flight diffractometer WISH at the ISIS  Facility in the UK. The sample was a 15~g powder of $\beta$-Na$_2$PrO$_3$ loaded in a thin-walled aluminium can. The powder contained a small amount of NaOH impurity, which resulted in an increased background signal. Diffraction patterns were collected at base temperature (1.4~K) and paramagnetic (10~K) for about 8~hours each at an average proton current of $30~\mu$A, with additional data collected at a selection of intermediate temperatures to obtain the order parameter. The raw time-of-flight neutron data were normalised and converted to $d$-spacing using the \textsc{mantid} \cite{Arnold2014} package. Rietveld refinements of crystal and magnetic structure models were performed using FullProf \cite{fullprof}, simultaneously against data measured in detector banks 1 to 10. A small absorption correction was included in the refinements to account for moderate neutron absorption by Na. The result of the structural refinement is presented in Supplementary Fig.~\swap{\ref{fig:neutron_nuclear}}{4}.

{\em Neutron spectroscopy.} Inelastic neutron scattering measurements were performed using the direct-geometry time-of-flight neutron spectrometer LET, also at ISIS. The sample was the same as for the powder neutron diffraction measurements described above. Most data was collected with LET operated in repetition rate multiplication mode to measure the inelastic scattering of neutrons with incident energies of $E_i\!=\! 3.7,7.5$ and 22.5~meV. The raw time-of-flight neutron data were corrected for detector efficiency and converted to intensity as a function of momentum transfer and energy $S(Q,\omega)$ using the \textsc{mantid} \cite{Arnold2014} package. The data in Fig.~\ref{fig:spinwave}a was counted for a total of 9~hours at an average proton current of $40~\mu$A. At higher energy transfers, the INS data showed visible non-dispersive inelastic peaks near 3 and 7.3~meV, attributed to well-known \cite{Holland} transitions between crystal-field levels of Pr$^{3+}$ ions in Pr$_6$O$_{11}$, present in the powder sample as a small impurity phase. This high-energy data was excluded from the analysis, at base temperature this signal is well isolated from the lower energy and strongly dispersive signal in Fig.~\ref{fig:spinwave}a attributed to cooperative magnetic excitations in the primary $\beta$-Na$_2$PrO$_3$ phase. Calculations of the spin-wave spectrum for model Hamiltonians were performed in the primitive cell with four magnetic sublattices using SpinW \cite{spinW}, for more details see Supplementary Notes 3-5. The spherically-averaged spin-wave spectrum was then compared to the experimentally measured $S(Q,\omega)$ to obtain the best fit Hamiltonian parameters.

\section*{Data availability} The experimental data supporting this research is openly available from ref.~\cite{data_archive}.

\section*{Acknowledgments} We acknowledge support from the European Research Council under the European Union’s Horizon 2020 research and innovation programme Grant Agreement Number 788814 (EQFT)(RO, KM, RC) and the Engineering and Physical Sciences Research Council (EPSRC) under grant No. EP/M020517/1(RC). RO acknowledges support from JSPS KAKENHI (Grant No. 23K19027JST) and JST ASPIRE (Grant No. JPMJAP2314). RC acknowledges support from the National Science Foundation under Grants No. NSF PHY-1748958 and PHY-2309135, and hospitality from the Kavli Institute for Theoretical Physics (KITP) where part of this work was completed. The neutron scattering measurements at the ISIS Facility were supported by beamtime allocations from the Science and Technology Facilities Council \cite{WISH_doi,LET_doi}. We thank Andrew Boothroyd for sharing his determination of the spherical magnetic form factor of Pr$^{4+}$ and for pointing out reference \cite{Holland} with crystal field transitions in Pr$_6$O$_{11}$.

\section*{Competing Interests}
The authors declare no competing interests.

\section*{Author Contributions}
RC, RDJ and RO conceived research, RO developed the synthesis protocol for powders and single crystals, and performed structural and magnetic characterization, KM performed single crystal heat capacity and torque experiments. RO, PM, RDJ and RC performed neutron powder diffraction measurements and RO analysed this data to solve the magnetic structure. RO, DV and RC performed inelastic neutron scattering measurements, RC and RO analysed this data and performed theoretical calculations. RO, RC and RDJ wrote the paper and the supplementary information with input from all co-authors. RC supervised all aspects of the project.

\bibliography{ref}
\clearpage

\begin{center}
\textbf{Supplementary Information}\\
\end{center}
Here we provide additional technical details on 1)~sample synthesis, 2)~refinement of the crystal structure from single-crystal x-ray and powder neutron diffraction, 3)~magnetic structure factor calculations and magnetic structure refinement, 4)~spin Hamiltonian for the hyperhoneycomb lattice, 5)~mean field description of the magnetic ground state depending on various terms in the spin Hamiltonian, and 6)~calculations of the spinwave spectrum and comparison with powder INS data.

\renewcommand{\thesection}{{Supplementary Note} \arabic{section}.\hspace{1mm}}
\renewcommand{\theequation}{\arabic{equation}} 
\renewcommand{\thetable}{\Roman{table}}
\renewcommand{\tablename}{\textbf{Supplementary Table}}
\renewcommand{\thefigure}{\arabic{figure}}
\renewcommand{\figurename}{\textbf{Supplementary Figure}}
\makeatother
\setcounter{figure}{0}
\setcounter{equation}{0}
\setcounter{section}{0}
\setcounter{table}{0}

\section{Synthesis}
\label{sec:synthesis}
\vspace{1.5mm}
Here we describe the powder synthesis of three polymorphs of Na$_2$PrO$_3$ and single crystal growth of $\beta$-Na$_2$PrO$_3$. Complementary synthesis studies focused mostly on growth of single crystals of $\alpha$-Na$_2$PrO$_3$ are provided in Ref.~\cite{InorganicChemistry}. All chemicals and samples were handled inside a nitrogen filled glovebox unless otherwise stated.

\textbf{Powder synthesis of three polymorphs of Na$_2$PrO$_3$.} Polycrystalline samples of $\alpha$- and $\beta$-Na$_2$PrO$_3$ were synthesized by annealing cubic-Na$_2$PrO$_3$. The cubic polymorph was first synthesized by a conventional solid-state reaction of Na$_2$O$_2$ (Alfa Aesar, 95$\%$) and Pr$_6$O$_{11}$ (Merck Life Science, 99.9$\%$). Pr$_6$O$_{11}$ was calcined in air at 800$^\circ$C for 24 hours. In a typical synthesis, 1.3 mmol of Pr$_6$O$_{11}$ and 8.6 mmol of Na$_2$O$_2$, which amounts to 10 mol\% excess use of Na$_2$O$_2$, were thoroughly ground and pressed into a pellet of $\phi$ = 12 mm in diameter. The pellet was loaded in an evacuated ($P<1$ Pa) 40 cm long $\phi$ = 17 mm diameter quartz tube. The ampoule was placed in a horizontal furnace and reacted at 400$^\circ$C for 48 hours. The heated sample contained purely the cubic phase, weighing 1.8 g. The heating time was determined such that all the excess Na$_2$O$_2$ is absorbed in the quartz ampouple after the reaction. The polycrystalline cubic phase sample was thoroughly ground, pressed into a pellet of $\phi$ = 12 mm in diameter, and loaded in an open silver tube inside the glovebox. The silver tube was sealed inside an evacuated quartz tube and reacted at 600$^\circ$C and 800$^\circ$C for 12 hours to obtain powder $\alpha$- and $\beta$- phases, respectively.

\textbf{Single crystal growth of $\beta$-Na$_2$PrO$_3$.} Single crystals of $\beta$-Na$_2$PrO$_3$ could be obtained by a solid-state reaction of Li$_8$PrO$_6$ and Na$_2$O as described by Wolf et al.~\cite{wolf19882}. Li$_8$PrO$_6$ was synthesized by heating stoichiometric mixture of as-received Pr$_6$O$_{11}$ (Merck Life Science, 99.9$\%$) and Li$_2$O (Alfa Aesar, 99.5$\%$) in oxygen flow at 700$^\circ$C for 24 hours. Li$_8$PrO$_6$ and Na$_2$O (Alfa Aesar, 80$\%$) were thoroughly ground and pressed into a pellet of $\phi$ = 5 mm in diameter. The pellet was placed in a silver tube of $\phi$ = 6 mm in diameter and then sealed by flame. After heating at 700$^\circ$C for several weeks, up to 0.5 mm single crystals were obtained.

\begin{table}
\caption{Fractional atomic coordinates and atomic displacement parameters of $\beta$-Na$_2$PrO$_3$ deduced from single-crystal x-ray diffraction at room temperature. Atomic coordinates, and equivalent isotropic $U_\text{eq}$ and anisotropic $U_{ij}$ displacement parameters (in units of 10$^{-3}$\AA$^2$) with estimated standard deviations in parentheses. $U_\text{eq}$ is defined as $(U_{11} + U_{22} + U_{33})/3$. Space group: $Fddd$ (origin choice 2 at $\bar{1}$), $a$ = 6.7641(2) \AA, $b$ = 9.7866(4) \AA, c = 20.5517(6) \AA, number of observed reflections = 4958, $R_\mathrm{int}$ = 5.88\%, $R(I>2\sigma(I))$ = 1.85\%, w$R(I>2\sigma(I))$ = 3.09\%, $S$=0.9180. Extinction corrections were applied by fitting the raw experimental structure factor magnitudes to
$|F_\text{calc}|/\left(1+0.001\xi|F_\text{calc}|^2\lambda^3/\sin2\theta\right)^{1/4}$, where $F_\text{calc}$ is the calculated structure factor, $\lambda=0.71073$~\AA~is the x-ray wavelength and $2\theta$ is the total scattering angle where the reflection is observed; $\xi$ was refined to 0.00080(3).}
\begin{ruledtabular}
\begin{tabular}{cccccc}
Site&
Wyckoff&
$x$&
$y$&
$z$&
$U_\text{eq}$\\
\colrule
Pr & $16g$ & 1/8 & 1/8 & 0.70879(1) & 6.4(1)\\
Na1& $16g$ & 1/8 & 1/8 & 0.0463(1) & 15.6(5)\\
Na2& $16g$ & 1/8 & 1/8 & 0.8796(1) & 11.3(5)\\
O1 & $16e$ & 0.8400(5) & 1/8 & 1/8 & 10.0(9)\\
O2 & $32h$ & 0.6384(5) & 0.3522(3) & 0.0336(1) & 9.7(6)\\
\colrule
$U_{11}$&$U_{22}$&$U_{33}$&$U_{12}$&$U_{13}$&$U_{23}$\\
\colrule
7.9(2) & 6.8(2) & 4.7(2) & 0.7(2) & 0 & 0\\
8(1) & 25(1) & 13(1) & 2(2) & 0 & 0\\
15(1) & 10(1) & 9(1) & 5(1) & 0 & 0\\
11(2) & 10(2) & 9(2) & 0 & 0 & 0.5(18)\\
10(2) & 11(1) & 9(1) & 1(2) & 0.5(12) & -0.9(11)\\
\end{tabular}
\end{ruledtabular}
\label{tab:XRD}
\end{table}

\section{Structural Characterization}
\label{sec:structure}
\vspace{1.5mm}
\textbf{Structural refinement from single-crystal x-ray and powder neutron diffraction.} Structural information obtained from refinement of single-crystal x-ray diffraction is presented in Supplementary Table~\ref{tab:XRD} and the quality of the agreements between data (top row) and refined model (middle row) is illustrated in Supplementary Fig.~\ref{fig:xray_exp_calc}. Note the observed diffraction patterns show very sharp peaks with no detectable diffuse scattering, as expected for a fully-ordered crystal structure, with no indication of structural stacking faults, in contrast to the case of powder samples of the layered polymorph $\alpha$-Na$_2$PrO$_3$ reported to have extensive layer stacking faults \cite{powderalphaNa2PrO3}. Although a lower symmetry, monoclinic $C2/c$ space group was originally proposed for $\beta$-Na$_2$PrO$_3$ in the original report in Ref.~\cite{wolf19882}, our extensive single crystal x-ray diffraction data shows that the higher-symmetry orthorhombic space group $Fddd$ can describe the intensities of all observed peaks just as well, yielding an $R_\mathrm{int}$ of 5.88\%, essentially indistinguishable from 5.74\% for $C2/c$. Furthermore, the $C2/c$ model predicts several additional diffraction peaks in the ($0kl$) and ($hk0$) planes that are not observed in the data, compare Supplementary Fig.~\ref{fig:xray_exp_calc} bottom and top rows. We therefore adopt the orthorhombic structure, which is isostructural to $\beta$-Li$_2$IrO$_3$ \cite{takayama2015hyperhoneycomb}. The structural refinement performed using \textsc{shelx} \cite{shelx} gives a fully-ordered structure, with no site mixing and nearly isotropic $U_{ij}$ with good agreement between the calculated and observed intensity indicated by small R factors (Supplementary Fig.~\ref{fig:xray_res}a). The crystal structure is schematically illustrated in Supplementary Fig.~\ref{fig:xray_res}b, all Pr sites (dark/light blue shaded spheres) are symmetry-equivalent, located inside three-fold coordinated edge-sharing O$_6$ octahedra, which form zigzag chains shown in dark/light blue along the $\bmm{a}\pm\bmm{b}$ diagonals.

Structural refinement of the neutron powder diffraction patterns at 10~K in the paramagnetic phase are shown in Supplementary Fig.~\ref{fig:neutron_nuclear} for detector banks that give access to different $d$-spacing ranges, with the resulting structural parameters listed in Supplementary Table~\ref{tab:neutron_nuclear_Na2PrO3}, the fractional coordinates are very similar with those obtained from the single-crystal x-ray refinement in Supplementary Table~\ref{tab:XRD}. Supplementary Tables~\ref{tab:neutron_nuclear_NaOH} and \ref{tab:neutron_nuclear_Pr6O11} show the results of the refinement of the NaOH and Pr$_6$O$_{11}$ impurity phases (2\% and 1\% weight phase fractions, respectively), present in the powder sample.
\begin{table}
\caption{\label{table:strucure_nd}Structural parameters of $\beta$-Na$_2$PrO$_3$ obtained from the refinement of powder neutron diffraction data at 10~K. Space group: $Fddd$, $a$ = 6.74560(12) \AA, $b$ = 9.74653(15) \AA, $c$ = 20.4972(4) \AA.}
\begin{ruledtabular}
\begin{tabular}{cccccc}
Site&
Wyckoff&
$x$&
$y$&
$z$&
10$^3\times U_\text{iso}$ (\AA$^2$)\\
\colrule
Pr & $16g$ & 1/8 & 1/8 & 0.7087(2) & 7.3(8)\\
Na1& $16g$ & 1/8 & 1/8 & 0.0455(3) & 11.7(7)\\
Na2& $16g$ & 1/8 & 1/8 & 0.8788(3) & 11.7(7)\\
O1 & $16e$ & 0.8422(5) & 1/8 & 1/8 & 0.4(4)\\
O2 & $32h$ & 0.6377(4) & 0.3547(1) & 0.03406(6) & 0.4(4)\\
\end{tabular}

\begin{tabular}{cccccc}
&
Bank 1&
Bank 2&
Bank 3&
Bank 4&
Bank 5\\
\colrule
$R_\mathrm{Bragg}$ & 9.16 & 4.17 & 3.67 & 4.03    &3.55\\
\end{tabular}
\end{ruledtabular}
\label{tab:neutron_nuclear_Na2PrO3}
\end{table}

\begin{figure*}
\includegraphics[width=17.5cm]{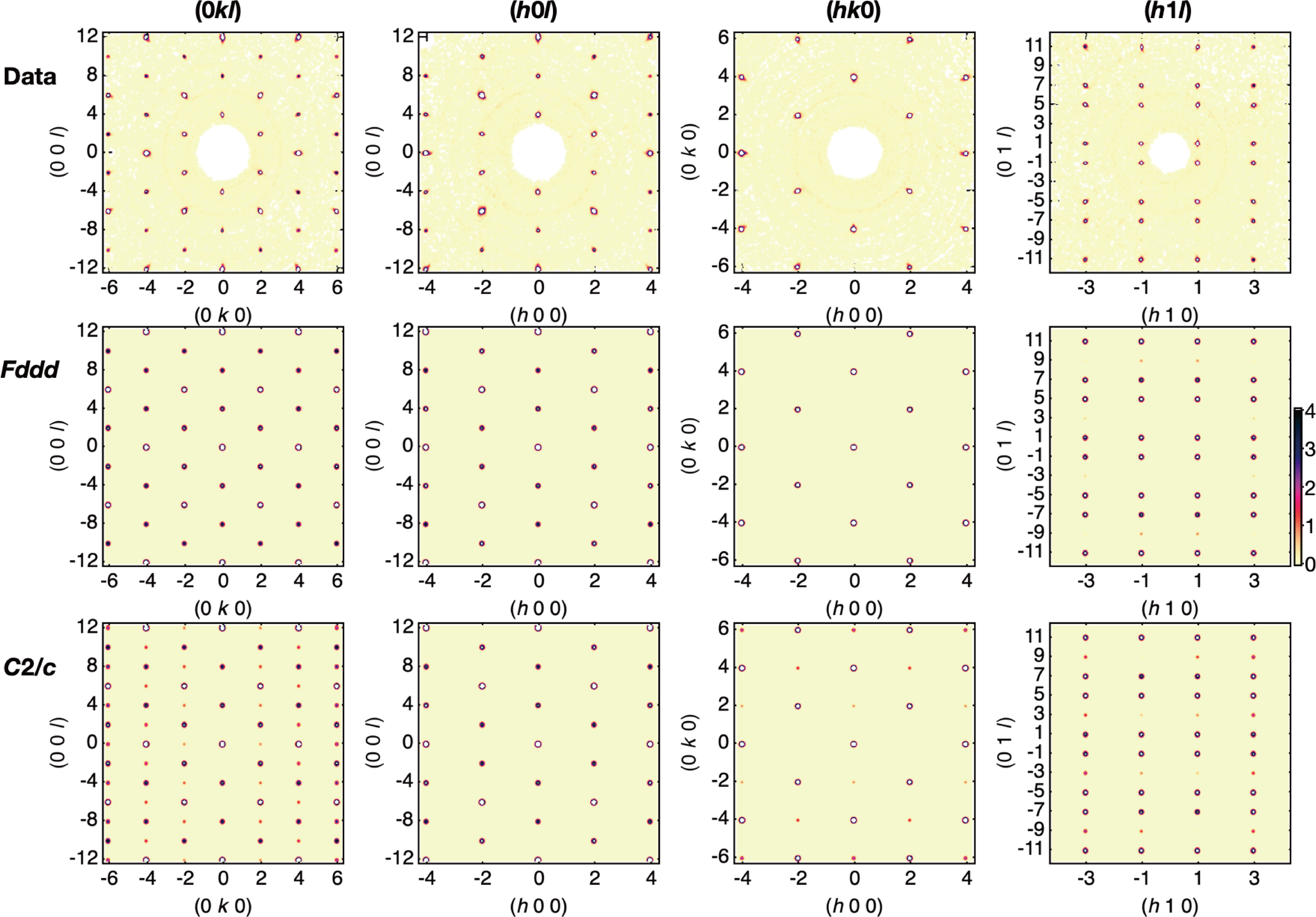}
\caption{\label{fig:xray_exp_calc} Single crystal x-ray diffraction patterns in representative planes, from left to right $0kl$, $h0l$, $hk0$ and $h1l$, indexed in r.l.u. units of the orthrohorhombic $Fddd$ unit cell. Top row shows the experimentally observed diffraction intensities, middle and bottom row are calculated patterns for the best-fit $Fddd$ structural model, and the $C2/c$ model proposed by Hoppe {\em et al.} \cite{wolf19882}, respectively. For the latter we used an idealised monoclinic unit cell with basis vectors (subscript $_m$) related to the orthorhombic $Fddd$ unit cell basis vectors via ${\bmm a}_\mathrm{m}= {\bmm a}$, ${\bmm b}_\mathrm{m}= -{\bmm b}$, ${\bmm c}_\mathrm{m}= -({\bmm c}+{\bmm a})/2$. In the $Fddd$ model, diamond glides perpendicular to the three orthorhombic axes allow only reflections with $h + k + l = 4p$ ($p$ integer) in the $0kl$, $h0l$ and $hk0$ planes. In contrast, the $C2/c$ model contains only one glide plane (perpendicular to the orthorhombic $b$-axis) and many peaks breaking the $h + k + l = 4p$ rule are predicted in the $0kl$ and $hk0$ planes (first and third columns), none of which are observed experimentally. In the $h1l$ plane (right-most column), because of face-centring, only $h$ odd, $l$ odd peaks are observed, and systematic weakening of the $l=\pm 3$ and $\pm 9$ peaks occurs because the crystal structure is nearly invariant by a translation by $\bmm{c}/6$ (for details see Supplementary Fig.~\ref{fig:c/6}). The colour bar indicates $\log_2(1+I/255)$, where $I$ is the diffraction intensity in arbitrary units.}
\end{figure*}

\begin{figure}[tbh]
\includegraphics[width=6.8cm]{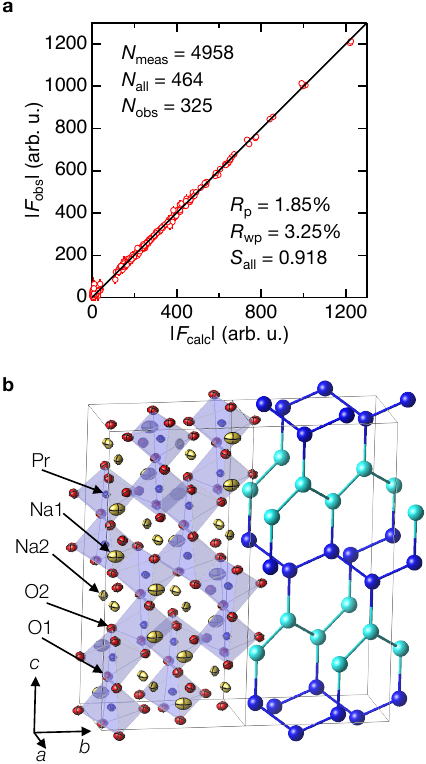}
\caption{\label{fig:xray_res}\textbf{a,} Observed versus calculated x-ray structure factor magnitudes for a single crystal. $|F_\text{obs}|$ is the experimental structure factor magnitude corrected for extinction as explained in Supplementary Table~{\ref{tab:XRD}. Error bars represent one standard deviation. Only reflections with $I>2\sigma(I)$ defined as observed peaks are included in the refinement, but all points with non-zero intensity are included in the plot}. Solid line shows the 1:1 agreement. A cluster of weak peaks $(|F_\mathrm{obs}|\! <\! 100)$ corresponds to all-odd peaks with $l = 3+6p$ or to all-even peaks with $h + k + l = 4p + 2$ ($p$ integer), which break a special reflection condition that applies to Na and Pr sites. \textbf{b,} Crystal structure obtained from the refinement of the single crystal XRD data with structural parameters listed in Supplementary Table~\ref{tab:XRD}. Blue/red/yellow ellipsoids in the left unit cell show the anisotropic displacement parameter ellipsoids of Pr, O, and Na atoms, respectively, with black line contours along the principal planes. Right unit cell shows the hyperhoneycomb lattice formed by Pr ions, with light/dark blue colours indicating the two families of zigzag chains running along the $\bmm{a}\mp\bmm{b}$ basal plane diagonals.}
\end{figure}

\begin{figure}[tbh]
\includegraphics{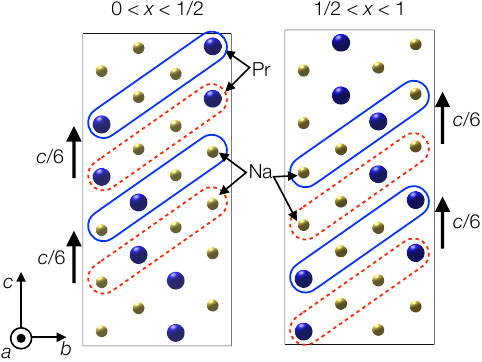}
\caption{\label{fig:c/6} Pseudo translational symmetry of the crystal structure of $\beta$-Na$_2$PrO$_3$. Projection of the atomic arrangement onto the $bc$ plane, left for $0<x<1/2$ and right for $1/2<x<1$. Small/large spheres represent Na/Pr, Oxygens are omitted for clarity. Dashed red/solid blue outlines indicate that the full atomic arrangement can be approximately reproduced starting with half the atoms (inside dashed red outlines) and translating them by approximately $\bmm{c}/6$ to obtain the other half of atoms (inside solid blue outlines). Destructive scattering interference between each atom and its translated pair leads to an almost exact cancellation of the structure factor for any ($hkl$) reflection with $l = 3+6p$ ($p$-integer). This approximate extinction rule becomes exact in the parent cubic NaCl structure.}
\end{figure}

\begin{figure*}
\includegraphics[width=15cm]{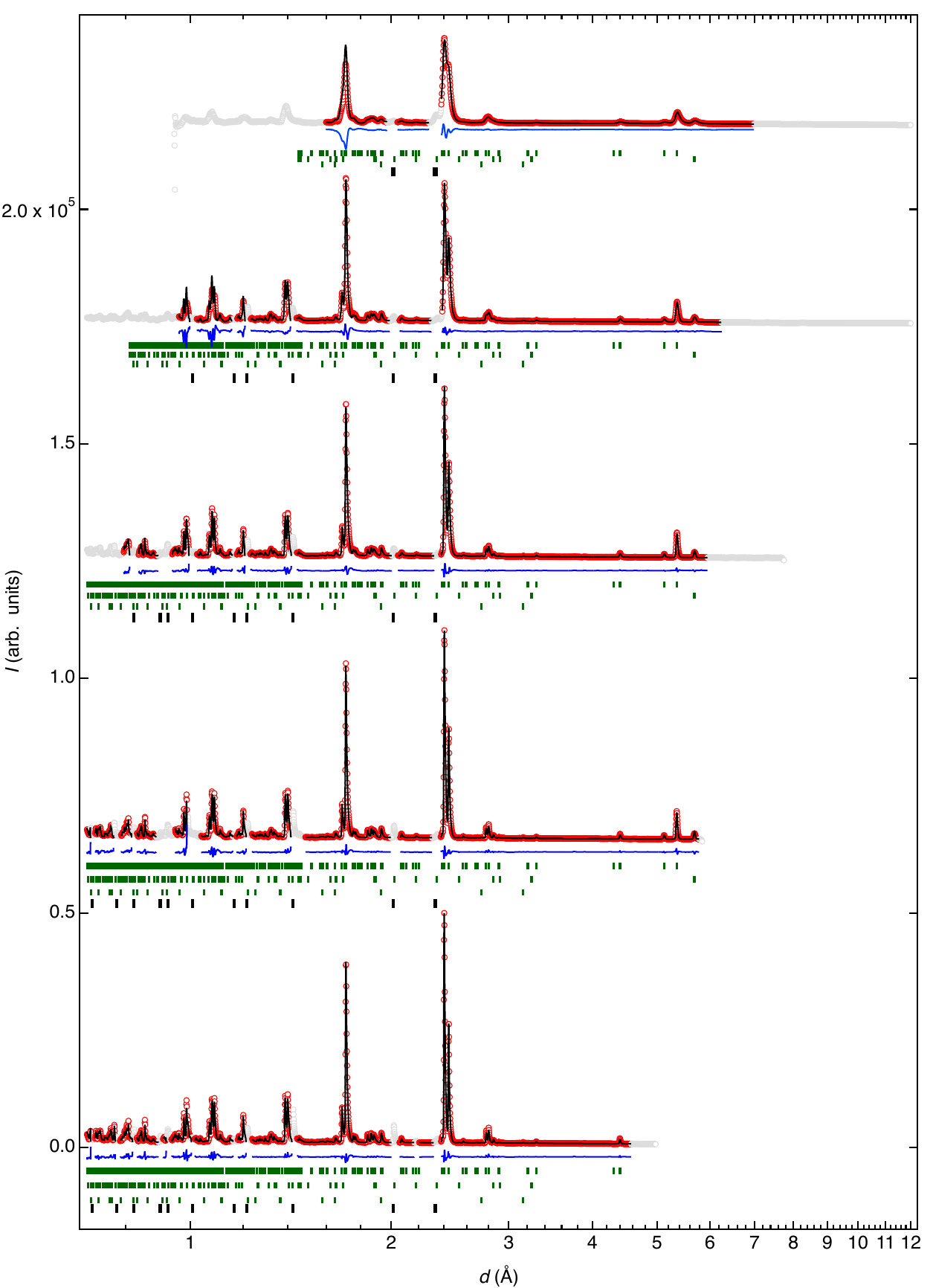}
\caption{\label{fig:neutron_nuclear} Structural refinement of the neutron powder diffraction patterns measured in the paramagnetic phase at 10~K. Data sets from detector banks with different $d$-spacing coverage are shown vertically offset for clarity with the horizontal axis the $d$-spacing on a log scale. Bank number increases from 1 (top) to 5 (bottom). Red open circles, black lines and blue lines indicate measured intensity, calculated intensity and residual of the fit, respectively. The rows of thin green bars below each pattern show the Bragg peak positions of the three refined phases, nominal $\beta$-Na$_2$PrO$_3$ (topmost row), NaOH (second row) and Pr$_6$O$_{11}$ (third row). The grayed out circles indicate data regions omitted from the fit, either because they are in close proximity of reflections from the aluminium sample holder (indicated by the bottom row of thick black bars under each graph), or they are in a low-resolution region of the pattern in some banks and higher-resolution data in the same $d$-spacing region in other banks is used instead. The agreement factors between the refinement in different banks are listed in Supplementary Table~\ref{tab:neutron_nuclear_Rfactors}.}
\end{figure*}

\textbf{Ideal crystal structure.} To make connection with theoretical models of spin Hamiltonians it is helpful to construct an ideal $Fddd$ structure with cubic PrO$_6$ octahedra. This is obtained by replacing the actual atomic fractional coordinates in Supplementary Table~\ref{tab:XRD} by ideal coordinates as follows: Pr $\left(\frac{1}{8}\frac{1}{8}\frac{17}{24}\right)$, Na1 $\left(\frac{1}{8}\frac{1}{8}\frac{1}{24}\right)$, Na2 $\left(\frac{1}{8}\frac{1}{8}\frac{7}{8}\right)$, O1 $\left(\frac{7}{8}\frac{1}{8}\frac{1}{8} \right)$, O2 $\left(\frac{5}{8}\frac{3}{8}\frac{1}{24}\right)$, and orthorhombic unit cell lattice parameters in ratio $a:b:c=1:\sqrt{2}:3$. In this ideal structure, by replacing Pr$\rightarrow$Na and O1,O2$\rightarrow$Cl, one recovers the $F$-centred cubic rock-salt NaCl structure oriented such that the orthorhombic unit cell axes are expressed in terms of the cubic cell axes as [202], [040], [$\bar{6}$06],  with $a=2\sqrt{2}a_0$, where $a_0$ is the cubic cell lattice parameter. Therefore, the actual crystal structure can be understood as a 2:1 Na:Pr cation ordering on the parent cubic rock-salt structure. The above relation to the parent cubic structure is at the origin of the (pseudo) translational symmetry of the actual $Fddd$ crystal structure by $\bmm{c}/6$ illustrated in Supplementary Fig.~\ref{fig:c/6}.

\begin{table}
\caption{Structural parameters of NaOH obtained from the refinement of powder neutron diffraction data at 10~K. Space group: $Bmmb$, $a$ = $b$ = 3.3854(6) \AA, $c$ = 11.377(1) \AA.}
\begin{ruledtabular}
\begin{tabular}{cccccc}
Site&
Wyckoff&
$x$&
$y$&
$z$&
10$^3\times U_\text{iso}$ (\AA$^2$)\\
\colrule
Na & $4c$ & 0 & 1/4 & 0.165(2) & 23(13)\\
O& $4c$ & 0 & 1/4 & 0.365(2) & 7(6)\\
H& $4c$ & 1/8 & 1/4 & 0.439(6) & 109(13)
\end{tabular}
\begin{tabular}{cccccc}
&
Bank 1&
Bank 2&
Bank 3&
Bank 4&
Bank 5\\
\colrule
$R_\mathrm{Bragg}$ &4.37 & 6.67 &9.19&  9.63& 10.8\\
\end{tabular}
\end{ruledtabular}
\label{tab:neutron_nuclear_NaOH}
\end{table}

\begin{table}
\caption{Structural parameters of Pr$_6$O$_{11}$ obtained from the refinement of powder neutron diffraction data at 10~K. Space group: $Fm\overline{3}m$, $a$ = $b$ = $c$ = 5.4567(8) \AA. Only the lattice constants are refined because of the insufficient number of observed peaks. The occupancies of Pr and O are fixed to 1 and 11/12, respectively, as expected from the chemical formula. Isotropic atomic displacement parameters of all the atoms are fixed to $0.5/8\pi^2=6.3\times10^{-3}$\AA$^2$.}
\begin{ruledtabular}
\begin{tabular}{ccccc}
Site&
Wyckoff&
$x$&
$y$&
$z$\\
\colrule
Pr & $4a$ & 1/4 & 1/4 & 1/4\\
O& $8c$ & 0 & 0 & 0\\
\end{tabular}
\begin{tabular}{cccccc}
&
Bank 1&
Bank 2&
Bank 3&
Bank 4&
Bank 5\\
\colrule
$R_\mathrm{Bragg}$ &27.3 & 13.7 & 13.6 & 10.8 & 10.2  \\
\end{tabular}
\end{ruledtabular}
\label{tab:neutron_nuclear_Pr6O11}
\end{table}

\begin{table}
\caption{Agreement factors of the structural refinement of powder neutron diffraction data at 10~K.}
\begin{ruledtabular}
\begin{tabular}{cccccc}
&
Bank 1&
Bank 2&
Bank 3&
Bank 4&
Bank 5\\
\colrule
$R_\mathrm{exp}$& 1.02 & 0.30 & 0.22 & 0.21 & 0.20\\
$R_\mathrm{p}$ & 15.5  & 11.8  & 8.97  & 9.04  & 8.88 \\
$R_\mathrm{wp}$& 9.37  & 6.64  & 6.94  & 7.28  & 7.74 \\
\end{tabular}
\end{ruledtabular}
\label{tab:neutron_nuclear_Rfactors}
\end{table}

\begin{table}
\caption{Fractional atomic coordinates of the Pr sites in the primitive cell, following the convention for site numbering defined in \cite{biffin2014unconventional} for isostructural $\beta$-Li$_2$IrO$_3$. Last four columns give the definition of the magnetic basis vectors $FCAG$ described in the text. Fractional atomic coordinates $xyz$ are with reference to the orthorhombic $Fddd$ unit cell and $z_\mathrm{Pr}$ = 0.7087(2) as described in Supplementary Table~\ref{table:strucure_nd}.}
\begin{ruledtabular}
\begin{tabular}{cccccccc}
\textrm{Pr site}&
\textit{x}&
\textit{y}&
\textit{z}&
\textit{F}&
\textit{C}&
\textit{A}&
\textit{G}\\
\colrule
$1$&1/8&1/8&$z_\mathrm{Pr}$&1&1&1&1\\
$2$&1/8&5/8&$3/4-z_\mathrm{Pr}$&1&1&-1&-1\\
$3$&3/8&3/8&$1-z_\mathrm{Pr}$&1&-1&-1&1\\
$4$&3/8&7/8&$1/4+z_\mathrm{Pr}$&1&-1&1&-1
\end{tabular}
\end{ruledtabular}
\label{tab:BV}
\end{table}

\begin{table}
\caption{Selection rules for the four magnetic basis vectors for all-even Miller indices and ideal Pr $z$-coordinate. In addition, all basis vectors contribute at all-odd indices.}
\begin{ruledtabular}
\begin{tabular}{cc}
\textrm{Basis Vector}&
\textrm{Reflection conditions}\\
\colrule
$F$&$l \neq 6p + 3,	h + k + l = 4p$\\
$C$&$l \neq 6p + 3,	h + k + l = 4p+2$\\
$A$&$l \neq 6p,	h + k + l = 4p$\\
$G$&$l \neq 6p,	h + k + l = 4p+2$
\end{tabular}
\end{ruledtabular}
\label{tab:reflection}
\end{table}

\begin{table}
\caption{Irreducible representations (irreps), basis vectors and magnetic space groups for $\bmm{q}\! =\! \bf{0}$ magnetic structures obtained using \textsc{isodistort} \cite{Stokes07}. Last two columns give the basis vectors of the magnetic unit cell and origin shift in terms of the lattice basis vectors of the structural cell, i.e. for the $m\Gamma^-_4$ irrep (last row) the $d'$ symbol in $Fd'dd$ refers to time reversal followed by a diamond glide normal to $\bmm{b}$, i.e. mirror in the ($x0z$) plane then translation by $\pm(\bmm{a} + \bmm{c})/4$.}
\begin{ruledtabular}
\begin{tabular}{ccccc}
\textrm{Irrep}&
\textrm{Basis}&
\textrm{Magnetic}&
\textrm{Unit}&
\textrm{Origin}\\
\textrm{}&
\textrm{Vectors}&
\textrm{Space Group}&
\textrm{Cell}&
\textrm{Shift}\\
\colrule
\noalign{\vskip 1pt}
$m\Gamma_1^+$&$G_z$&$Fddd.1$&$(\bmm{a},\bmm{c},-\bmm{b})$&$(1/4,3/2,1/4)$\\
$m\Gamma_2^+$&$F_z$&$Fd'd'd$&$(-\bmm{a}, -\bmm{b}, \bmm{c})$&$(7/4,7/4,0)$\\
$m\Gamma_3^+$&$F_x,G_y$&$Fd'd'd$&$(-\bmm{b}, -\bmm{c}, \bmm{a})$&$(0,7/4,7/4)$\\
$m\Gamma_4^+$&$G_x,F_y$&$Fd'd'd$&$(-\bmm{a}, -\bmm{c}, -\bmm{b})$&$(3/2,3/2,3/2)$\\
$m\Gamma_1^-$&$A_z$&$Fd'd'd'$&$(\bmm{a}, \bmm{c}, -\bmm{b})$&$(1/4,3/2,1/4)$\\
$m\Gamma_2^-$&$C_z$&$Fd'dd$&$(-\bmm{c}, -\bmm{b}, -\bmm{a})$&$(3/2,3/2,3/2)$\\
$m\Gamma_3^-$&$C_x,A_y$&$Fd'dd$&$(-\bmm{a}, -\bmm{c}, -\bmm{b})$&$(3/2,3/2,3/2)$\\
$m\Gamma_4^-$&$A_x,C_y$&$Fd'dd$&$(\bmm{b}, \bmm{c}, \bmm{a})$&$(0,0,0)$
\end{tabular}
\end{ruledtabular}
\label{tab:MSG}
\end{table}

\begin{table}
\caption{Structure factors for the four magnetic basis vectors, for all-odd or all-even reflections (for the ideal Pr $z$-coordinate).}
\begin{ruledtabular}
\begin{tabular}{cc}
Basis Vector&
Structure Factor $\mathcal{F}(hkl)$\\
\colrule
\noalign{\vskip 1pt}
$F$&$16(-1)^l\cos\left(\pi l/6\right)\cos\left[\pi(h+k+l)/4\right]$\\
$C$&$16i(-1)^l\cos\left(\pi l/6\right)\sin\left[\pi(h+k+l)/4\right]$\\
$A$&$16i(-1)^l\sin\left(\pi l/6\right)\cos\left[\pi(h+k+l)/4\right]$\\
$G$&$-16(-1)^l\sin\left(\pi l/6\right)\sin\left[\pi(h+k+l)/4\right]$\\
\end{tabular}
\end{ruledtabular}
\label{tab:SF}
\end{table}

\newpage
\section{Magnetic structure factors}
\label{sec:structure_factors}
{\bf Magnetic structure factor.} The magnetic structure factor for a magnetic Bragg peak at wavevector $\bmm{Q}$ is
\begin{equation}
\bm{\mathcal{F}}(\bmm{Q})=f_\text{F} \sum_n \bmm{m}_n e^{i\bmm{Q}\cdot\bmm{r}_n},
\label{eq:FQ}
\end{equation}
where the prefactor $f_\text{F}=1+e^{i\pi(h+k)}+e^{i\pi(k+l)}+e^{i\pi(l+h)}$ is due to the $F$-centering of the orthorhombic structural cell. The sum extends over all sites in the primitive unit cell ($n=1-4$), where $\bmm{m}_n$ is the magnetic moment at site $n$ located at position $\bmm{r}_n$. Each of the four magnetic basis vectors has symmetry-imposed relative orientations between the $\bmm{m}_{1-4}$ moments as listed in Supplementary Table~\ref{tab:BV}, i.e. for an $A$-basis vector $\bmm{m}_1=-\bmm{m}_2=-\bmm{m}_3=\bmm{m}_4$. Consider a magnetic structure in the basis vector combination ($A_x,\pm C_y$)  where the upper/lower sign corresponds to in-phase/out-of-phase relation between the two basis vectors, with $M_x$ and $M_y$ moment magnitudes ($M_{x,y}\!>\!0$) along the $x$ and $y$ directions for the magnetic moment at each site. The magnetic structure factor vector in this case is $\bm{\mathcal{F}}=M_x\mathcal{F}^A\bmm{\hat{a}} \pm M_y\mathcal{F}^C\bmm{\hat{b}}$, where $\mathcal{F}^A$ and $\mathcal{F}^C$ are the structure factors of the $A$ and $C$ basis vectors given in Supplementary Table~\ref{tab:SF}.

\begin{figure*}
\includegraphics[width=17.5cm]{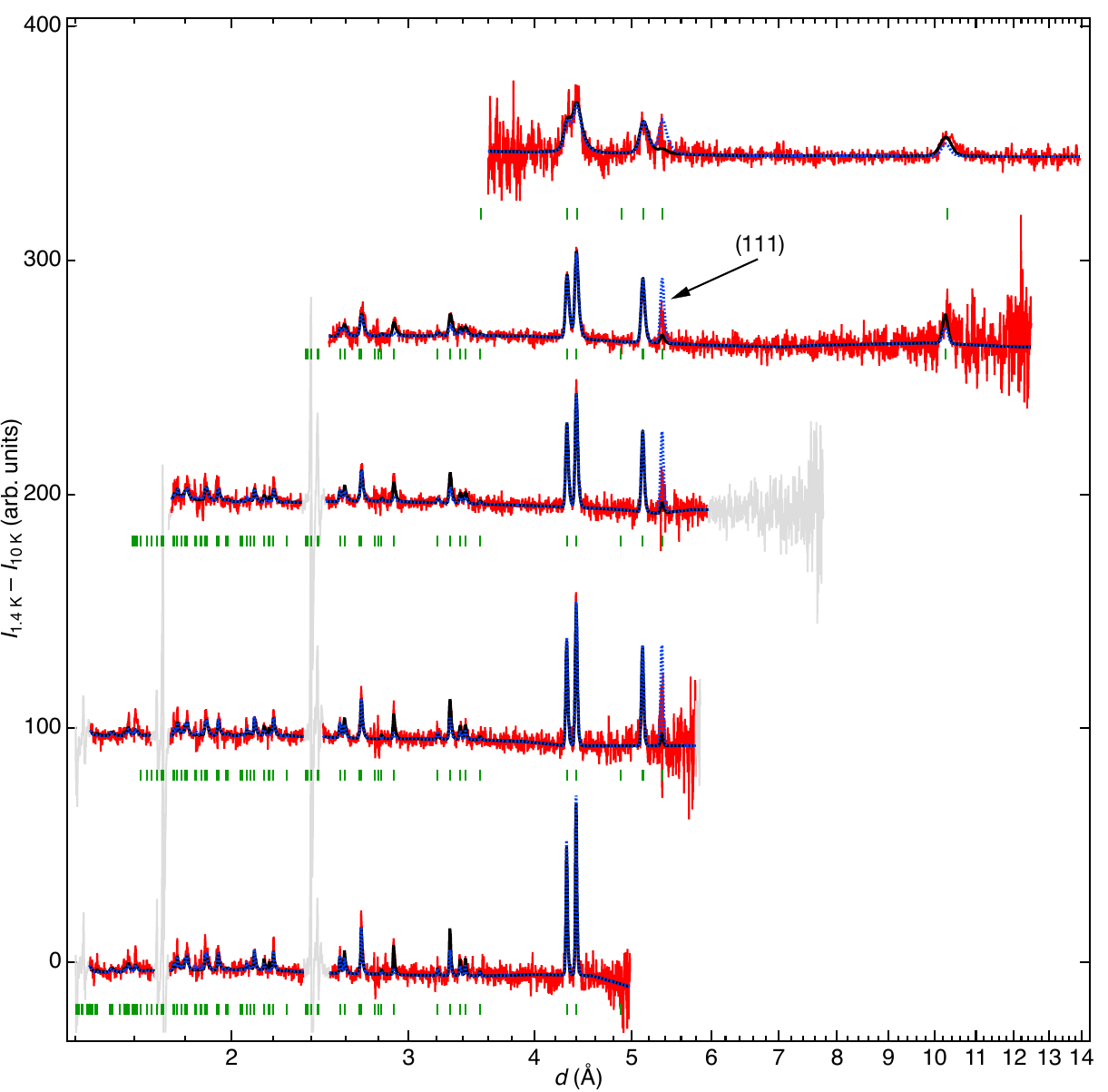}
\caption{\label{fig:neutron_magnetism} Magnetic neutron powder diffraction pattern at 1.4~K obtained after subtracting from the raw data the pattern measured in the paramagnetic phase at 10~K. The vertically-offset five data sets are for the same banks as in Supplementary Fig.~\ref{fig:neutron_nuclear} with horizontal axis the $d$-spacing on a log scale. Red lines and green bars indicate the subtracted intensity and Bragg peak positions, respectively. Gray shading indicates noisy regions removed from the fit. Black/blue dotted lines are best fits to the magnetic structure models ($A_x,\mp C_y$) described in the text. We note that the (111) peak near 5.4~\AA$\,\,$ is rather noisy because of a rather strong structural Bragg peak intensity subtracted off, so the signal there cannot be used to reliably discriminate between models. Data in all banks was refined simultaneously, the intensity scale factor in bank 2 was fixed to the value obtained from refining the structural pattern in that bank and intensity scale factors in the other banks were refined and values varied within $\sim\!15\%$ between banks. Bank 2 was chosen for the intensity normalization as the structural diffraction pattern in that bank could be fitted most accurately.}
\end{figure*}

{\bf Magnetic diffraction intensity.} The intensity of magnetic Bragg peaks observed in unpolarised neutron diffraction is proportional to the modulus squared of the magnetic structure factor vector perpendicular to the scattering wavevector $\bmm{Q}$, i.e. \begin{equation}I(\bmm{Q})=\left(\bm{\mathcal{F}}(\bmm{Q})\times\bmm{\hat{Q}}\right)\cdot\left(\bm{\mathcal{F}}^*(\bmm{Q})\times\bmm{\hat{Q}}\right),\label{eq:IQ}\end{equation}
where $^*$ indicates complex conjugation and $\bmm{\hat{Q}}$ is the unit vector along $\bmm{Q}$. Expanding the above expression gives
\begin{eqnarray}
I(\bmm{Q})&=&\left(1-\frac{Q^2_x}{Q^2}\right)M_x^2|\mathcal{F}^A|^2+\left(1-\frac{Q_y^2}{Q^2}\right)M_y^2|\mathcal{F}^C|^2 \nonumber \\ & & \mp~2M_xM_y\frac{Q_xQ_y}{Q^2}\text{Re}\left(\mathcal{F}^A\,\mathcal{F}^{C*}\right),
\label{eq:IofQ}
\end{eqnarray}
where $\text{Re}()$ indicates real part. The first two terms are the respective contributions of each of the two basis vectors separately and the last term is a cross-term that is directly sensitive to the relative sign between the two basis vectors. Magnetic Bragg peaks such as (131) can be used to discriminate between the two magnetic structure models ($A_x,\pm C_y$) as the cross-term is relatively large, obtained by direct calculation as $\mathcal{F}^A\mathcal{F}^{C*}=32\sqrt{3}>0$. Selecting the upper/lower sign in front of the intensity cross-term results in a lower/higher peak intensity, the difference is significant as illustrated by the solid black (lower sign)/dotted blue (upper sign) lines in Fig.~\ref{fig:diffraction}f (left side), the lower sign ($A_x,- C_y$) needs to be selected to correctly reproduce the observed peak intensity.

\begin{figure*}[tb]
\includegraphics[width=17.5cm]{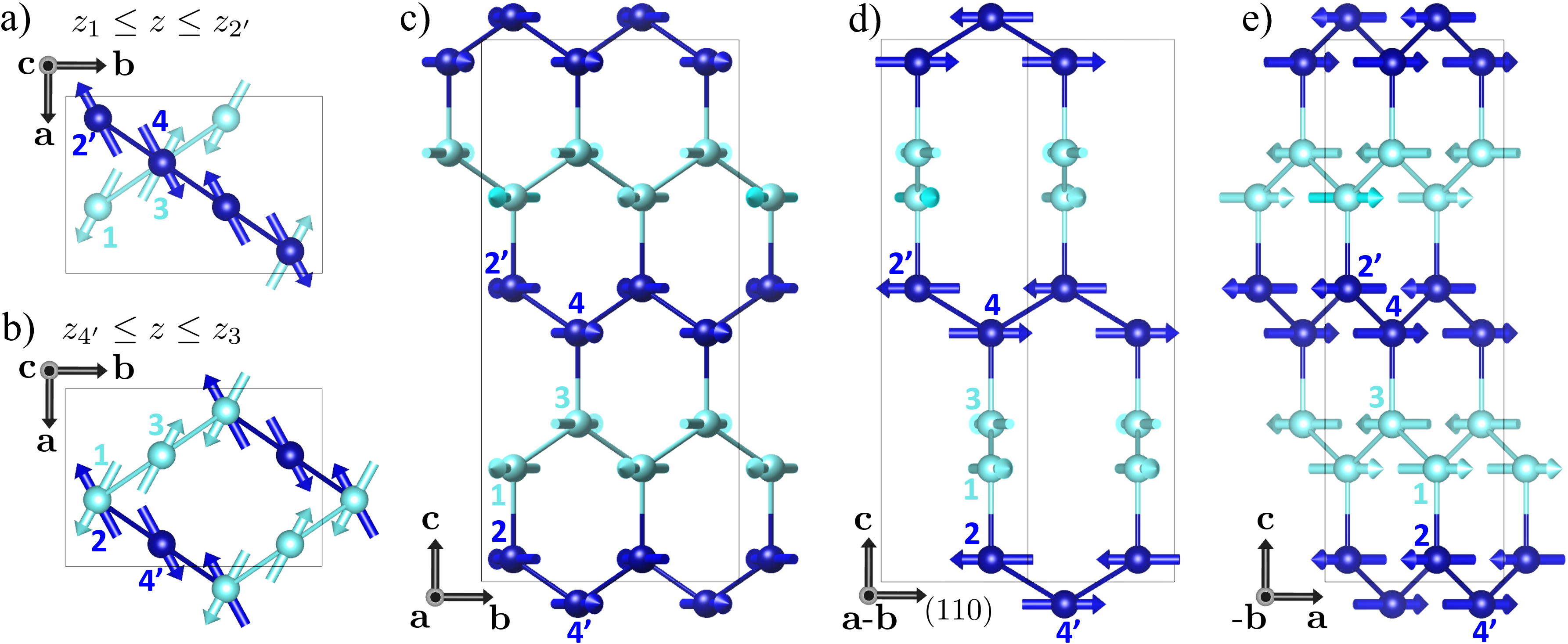}
\caption{\label{fig:magnetic_structure} Magnetic structure projected onto different crystallographic planes, from left to right $ab$, $bc$, $(110)\times\bmm{c}$ and $ac$. Dark/light blue balls (arrows) indicate Pr atoms (moments) on the two families of zigzag chains running along the basal plane diagonals $\bmm{a}\pm\bmm{b}$, respectively. Labels 1-4 are the same as in Fig.~\ref{fig:diffraction}g) and indicate sites equivalent to those in the primitive unit cell. $2'$ and $4'$ label sites obtained from 2 and 4 via the $F$-centring translation $\pm[-1/2,0,1/2]$, respectively. \textbf{a,b,} Projection of the magnetic structure onto the $ab$ plane for Pr atoms with $z$-coordinate in the range $[z_1,z_{2'}]$ and $[z_{4'},z_3]$, respectively, where $z_i$'s  are the $z$-coordinates of sites $i$ defined above. \textbf{d,} View of the magnetic structure along the $\bmm{a}-\bmm{b}$ direction of the light blue zigzag chains, which emphasises that magnetic moments are oriented almost along the direction of the chain they belong to, with the two families of chains making a large angle between them as illustrated in panels \textbf{a,b}. The magnetic structure has eight primary symmetry operations $\left\{{1,\bar{1}',2_y,2'_z,2'_x,d'_y,d_z,d_x}\right\}$, with $\bar{1}'$ located at the middle of every zigzag bond and all 2-fold axes passing through the middle of a vertical bond. $d'_y$ is time reversal followed by a diamond glide normal to $\bmm{b}$, i.e. mirror in ($x0z$) plane then translation by $\pm(\bmm{a}+\bmm{c})/4$, and the other diamond glides also pass through the origin. The VESTA file used to plot the magnetic structure is provided in Ref.~\cite{data_archive}.}
\end{figure*}

{\bf Effects of powder averaging.} In a powder diffraction experiment all magnetic Bragg peaks at wavevectors $\bmm{Q}$ related by symmetry operations of the lattice point group overlap. We have explicitly checked that all symmetry operations of the $mmm$ lattice point group will leave the intensity expression $I(\bmm{Q})$ invariant for all-odd reflections where both $A$ and $C$ basis vectors contribute. This means that all averaged peaks in the powder pattern have the same intensity and the same cross-term, so the powder diffraction data is just as sensitive to the phase difference between the two basis vectors as a single crystal experiment would be.

{\bf Symmetry of the magnetic diffraction pattern in reciprocal space.} The invariance of the intensity expression $I(\bmm{Q})$ under symmetry operations of the $mmm$ lattice point group can also be deduced using more general symmetry considerations. The magnetic space group of the experimentally-determined magnetic structure illustrated in Supplementary Fig.~\ref{fig:magnetic_structure} is $Fd'dd$ with unit cell basis vectors ($\bmm{b}$,$\bmm{c}$,$\bmm{a}$), with eight primary symmetry operations $\left\{{1,\bar{1}',2_y,2'_z,2'_x,d'_y,d_z,d_x}\right\}$, and their $F$-centred versions, where $xyz$ are along the $abc$ axes of the structural cell. Here $'$ indicates time reversal, $\bar{1}'$ is located at the origin and the middle of every zigzag bond, all 2-fold axes pass through the middle of a vertical bond such as $\left(\frac{3}{8},\frac{3}{8},\frac{3}{8}\right)$, $d'_y$ is time reversal followed by a diamond glide normal to $\bmm{b}$, i.e. mirror in ($x0z$) plane then translation by $\pm(\bmm{a}+\bmm{c})/4$, and the other diamond glides also pass through the origin. The corresponding magnetic point group is $m'mm$ with symmetry operations $\left\{{1,\bar{1}',2_y,2'_z,2'_x,m'_y,m_z,m_x}\right\}$. $I(\bmm{Q})$ is therefore invariant under those point group operations. Furthermore, one can see by inspection of eqs.~(\ref{eq:FQ},\ref{eq:IQ}) that $I(\bmm{Q})$ is also invariant under inversion $\bar{1}$, which maps $\bmm{Q}\rightarrow -\bmm{Q}$ and $\bm{\mathcal{F}}(\bmm{Q})\rightarrow - \bm{\mathcal{F}}(-\bmm{Q})=-\bm{\mathcal{F}}^*(\bmm{Q})$ (using that $\bmm{m}_{1-4}$ are real), so combining inversion with the above point group operations gives that $I(\bmm{Q})$ is invariant under all symmetry operations of the paramagnetic point group $mmm1'$, which contains all operations of the $mmm$ lattice point group. Therefore all powder-averaged magnetic reflections have the same intensity.

{\bf Magnetic domains.} Since the magnetic point group $m'mm$ has half the symmetry operations of the paramagnetic point group $mmm1'$, there can be two magnetic domains, related by time reversal. Time-reversed domains have identical diffraction patterns, as time-reversal maps $\bmm{Q}\rightarrow -\bmm{Q}$ and $\bm{\mathcal{F}} (\bmm{Q})\rightarrow - \bm{\mathcal{F}}(-\bmm{Q})=-\bm{\mathcal{F}}^*(\bmm{Q})$.

{\bf Magnetic structure refinement.} Results of the refinement of the magnetic neutron diffraction pattern, obtained from the raw low temperature (1.4~K) pattern by subtracting the paramagnetic (10~K) pattern, is shown in Supplementary Fig.~\ref{fig:neutron_magnetism}. Good consistency is obtained between the detector banks probing different $d$-spacing ranges.

\begin{table}
\caption{Agreement factors of the magnetic refinement of the powder neutron diffraction data (1.4~K) to the ($A_x, -C_y$) magnetic structure model. The refined moment sizes (in $\mu_\text{B}$) are $M_x = 0.195(3)$ and $M_y = 0.107(4)$.}
\begin{ruledtabular}
\begin{tabular}{cccccc}
&
Bank 1&
Bank 2&
Bank 3&
Bank 4&
Bank 5\\
\colrule
$R_\mathrm{exp}$& 84.6 & 55.3 & 54.8 & 62.7 & 66.0 \\
$R_\mathrm{p}$ & 236 & 204 & 125 & 158 & 158\\
$R_\mathrm{wp}$& 71.8 & 48 & 49.2 & 56.8 & 60.1\\
$R_\mathrm{Bragg}$& 18.8 & 24.3 & 16.1 & 22.5 & 29.3\\
\end{tabular}
\end{ruledtabular}
\label{tab:neutron_magnetic_Rfactors_AxminusCy}
\end{table}

\begin{table}
\caption{Agreement factors of the magnetic refinement of the powder neutron diffraction data (1.4~K) based to the ($A_x, C_y$) magnetic structure model. The refined moment sizes (in $\mu_\text{B}$) are $M_x = 0.192(3)$ and $M_y = 0.079(4)$.}
\begin{ruledtabular}
\begin{tabular}{cccccc}
&
Bank 1&
Bank 2&
Bank 3&
Bank 4&
Bank 5\\
\colrule
$R_\mathrm{exp}$& 84.6  & 55.3  & 54.8 & 62.7  & 66\\
$R_\mathrm{p}$ & 246& 211& 136& 165& 163\\
$R_\mathrm{wp}$& 77.1  & 50.8  & 53.1  & 59.4  & 63.1 \\
$R_\mathrm{Bragg}$& 39.6 & 41.9 & 34.8 & 39.3 & 35.4\\
\end{tabular}
\end{ruledtabular}
\label{tab:neutron_magnetic_Rfactors_AxplusCy}
\end{table}

\section{Spin Hamiltonian}
\label{sec:Ham}
\vspace{1.5mm}
{\bf Definition of cubic $\sf{xyz}$ axes.} For discussing the magnetic exchange it is convenient to use as reference the ideal crystal structure with cubic PrO$_6$ octahedra described in Supplementary Note 2. In this case normals to the three Pr-O$_2$-Pr superexchange planes meeting at a site are reciprocally orthogonal and define a cubic axes frame $\sf{xyz}$ (SansSerif font to distinguish them from the orthorhombic $xyz$ axes), with each Pr-Pr bond colour coded red/green/blue according to the axis normal to its superexchange plane. This is illustrated in Fig.~\ref{fig:structure}d, where the top triad of axes shows how the cubic axes are related to the orthorhombic axes of the ideal structure following the convention introduced in Ref.~\cite{lee2015hyperhoneycomb}, namely
$$\left(\begin{array}{c} \bf{\hat{\sf{x}}}\\ \hat{\sf{y}} \\ \hat{\sf{z}}\end{array}\right)
=\mathcal{R} \left(\begin{array}{c} \bmm{\hat{a}}\\ \bmm{\hat{b}} \\ \bmm{\hat{c}}\end{array}\right) \quad \text{with} \quad
\mathcal{R}=\left(\begin{array}{ccc} -\frac{1}{\sqrt{2}} & 0 & -\frac{1}{\sqrt{2}} \\ \frac{1}{\sqrt{2}} & 0 & -\frac{1}{\sqrt{2}} \\ 0 & -1 & 0 \end{array}\right).$$

\begin{figure}
\includegraphics[width=7cm]{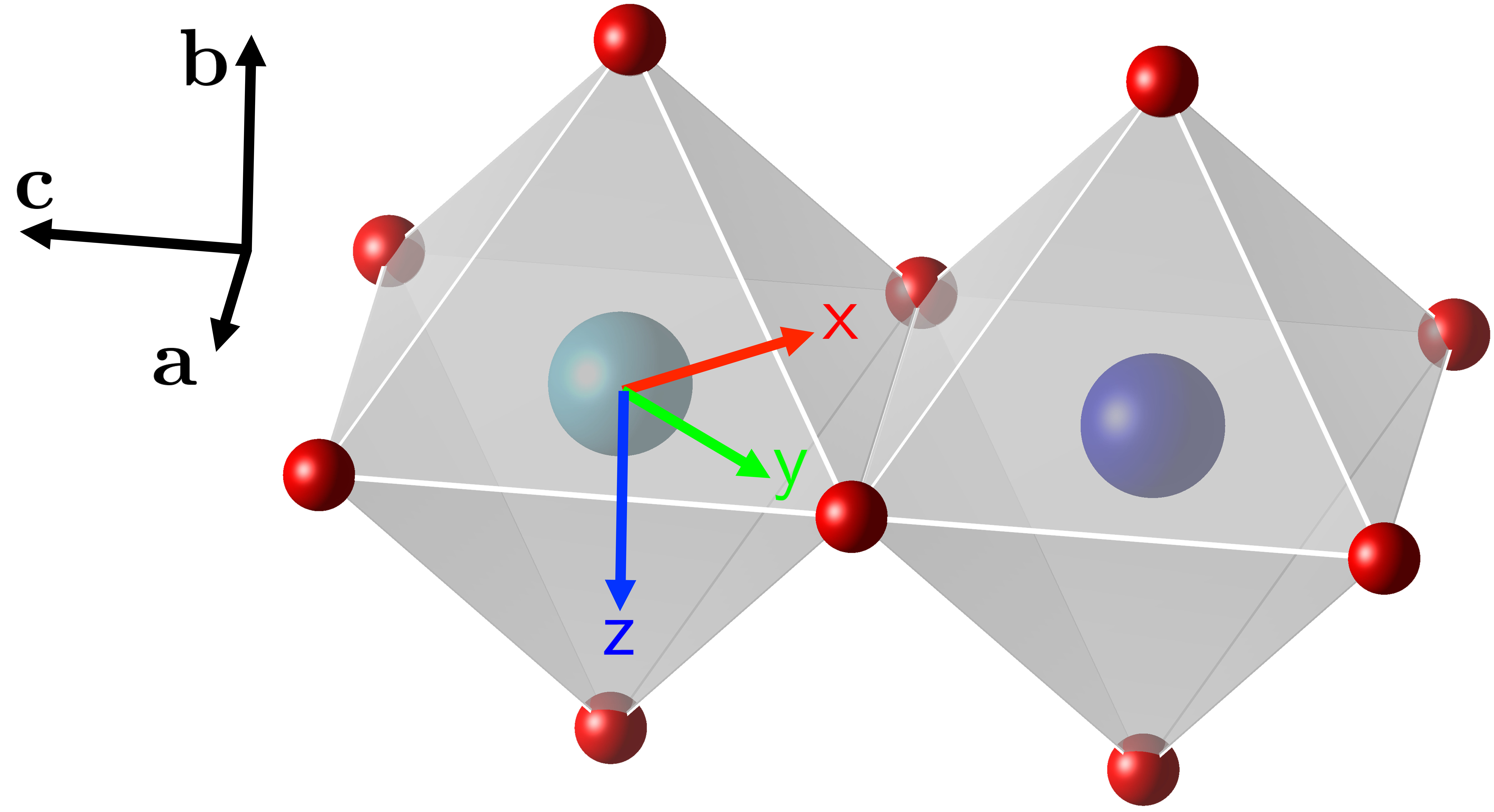}
\caption{\label{fig:zbond} Edge-sharing PrO$_6$ octahedra for a $\sf{z}$ bond in the ideal crystal structure showing the orientation of the $\sf{xyz}$ axes relevant to discuss the exchange Hamiltonian: $\sf{z}$ is normal to the Pr-O$_2$-Pr superexchange plane and $\sf{x}$ and $\sf{y}$ are along the Pr-O bonds in this plane (light/dark blue spheres are Pr, red balls are O).}
\end{figure}

{\bf Exchange for $\sf{z}$-bonds.} For a $\sf{z}$-bond as illustrated in Supplementary Fig.~\ref{fig:zbond}, the $JK\Gamma$ Hamiltonian has the form $\mathcal{H}_{ij}^{\sf{z}}=J\bmm{S}_i\cdot\bmm{S}_j + K S_i^{\sf{z}}S_j^{\sf{z}} + \Gamma(S_i^{\sf{x}}S_j^{\sf{y}}+S_i^{\sf{y}}S_j^{\sf{x}})$ where $i$ and $j$ index the two spin sites at the ends of the bond. In matrix notation, $\mathcal{H}_{ij}^{\sf{z}}=\bmm{S}_i \mathcal{J}^{\sf z} \bmm{S}_j^T$, where $\bmm{S}_i$ is a shorthand notation for the row vector of spin components $[S^{\sf{x}}_i \, S^{\sf{y}}_i \, S^{\sf{z}}_i]$, $^T$ is matrix transpose, and the exchange matrix is \begin{equation}\mathcal{J}^{\sf z}=\left(\begin{array}{ccc} J & \Gamma & 0 \\ \Gamma & J & 0 \\ 0 & 0 & J+K \end{array}\right).\label{eq:Jz_xyz}\end{equation} In orthorhombic axes (subscript $_{abc}$) the exchange matrix is obtained as \begin{equation}\mathcal{J}^{\sf z}_{abc}=\mathcal{R}^T\mathcal{J}^{\sf z}\mathcal{R}=\left(\begin{array}{ccc} J-\Gamma & 0 & 0 \\ 0 & J+K & 0 \\ 0 & 0 & J+\Gamma \end{array}\right).\label{eq:Jz}\end{equation} All (blue) $\sf{z}$-bonds are symmetry-equivalent via $F$-centring translations or inversion centres located in the middle of every zigzag bond.

{\bf Energetic selection of spin orientations on $\sf{z}$-bonds.} Considering for simplicity the case of $K=0$, $J>0$ aligns the spins at the ends of the bond in a collinear antiparallel arrangement, and the effect of a finite $\Gamma$ is to introduce an energy dependence on orientation. For $\Gamma<0$ the directions in order of increasing energy are $\hat{\sf{x}}-\hat{\sf{y}}$ ($\parallel a$), $\hat{\sf{z}}$ ($\parallel b$), $\hat{\sf{x}}+\hat{\sf{y}}$ ($\parallel c$), i.e. the energetically-preferred direction is in the exchange plane normal to the bond direction. On the other hand, for $\Gamma>0$ the directions in order of increasing energy are $\hat{\sf{x}}+\hat{\sf{y}}$, $\hat{\sf{z}}$, $\hat{\sf{x}}-\hat{\sf{y}}$, i.e. the energetically-preferred direction is along the bond direction.

{\bf Exchange for $\sf{x}$-bonds.} To discuss the exchange on $\sf{x}$-bonds, we choose the representative (red) 1-3 bond on a $\bmm{a}-\bmm{b}$ chain in Fig.~\ref{fig:diffraction}g. The $JK\Gamma$ Hamiltonian for this bond has the form $\mathcal{H}_{ij}^{\sf{x}}=J'~\bmm{S}_i~\cdot~\bmm{S}_j + K'~S_i^{\sf{x}}~S_j^{\sf{x}} + \Gamma'~(S_i^{\sf{y}}S_j^{\sf{z}}+S_i^{\sf{z}}S_j^{\sf{y}})$ where $i$ and $j$ index the two spin sites at the ends of the bond. $\sf{x}$- and $\sf{z}$-bonds are symmetry-inequivalent, i.e. there is no symmetry operation of the crystal space group that maps one onto the other. The theoretical study of Ref.~\cite{lee2015hyperhoneycomb} considered the magnetic phase diagram under the simplifying assumption that $\sf{x}$- and $\sf{z}$-bonds are related by a (pseudo) 3-fold rotation around the axis normal to the plane defined by the two bonds, which however is not a symmetry operation even for the ideal structure described in \ref{sec:structure} For this simplified case $J'=J$, $K'=K$ and $\Gamma'=\Gamma$, which can be seen with reference to Fig.~\ref{fig:diffraction}g: the (red) 1-3 $\sf{x}$-bond is obtained from the (blue) 1-2 $\sf{z}$-bond via rotation by +$120^{\circ}$ around $\bmm{\hat{n}}_1=(-\hat{\sf{x}}+\hat{\sf{y}}-\hat{\sf{z}})/\sqrt{3}$, which maps $\hat{\sf{x}}\rightarrow-\hat{\sf{y}}$, $\hat{\sf{y}}\rightarrow-\hat{\sf{z}}$ and $\hat{\sf{z}}\rightarrow\hat{\sf{x}}$, therefore mapping $\mathcal{H}^{\sf{z}}_{12} \rightarrow\mathcal{H}^{\sf{x}}_{13}$. However, in the following, we treat the $\sf{x}$- and $\sf{z}$-bonds as symmetry-inequivalent, unless explicitly stated otherwise.

In matrix notation, in the two frames the exchange matrix for the representative 1-3 $\sf{x}$-bond is \begin{equation}\mathcal{J}^{\sf x}=\left(\begin{array}{ccc} J'+K' & 0 & 0 \\ 0 & J' & \Gamma' \\ 0 & \Gamma' & J' \end{array}\right)\label{eq:Jx_xyz}\end{equation} with
\begin{equation}\mathcal{J}^{\sf x}_{abc}=\left(\begin{array}{ccc} J'+\frac{K'}{2} & -\frac{\Gamma'}{\sqrt{2}} & \frac{K'}{2} \\ -\frac{\Gamma'}{\sqrt{2}} & J' & \frac{\Gamma'}{\sqrt{2}} \\ \frac{K'}{2} & \frac{\Gamma'}{\sqrt{2}} & J'+\frac{K'}{2} \end{array}\right).\label{eq:Jx}\end{equation}

{\bf Exchange for $\sf{y}$-bonds.} The (green) $\sf{y}$- and (red) $\sf{x}$-bonds are symmetry equivalent. For example the $\sf{y}$- and $\sf{x}$-bonds sharing a common site are related by a 2-fold rotation along the $\sf{z}$-bond sharing the same site, this rotation maps $\hat{\sf x} \rightarrow \hat{\sf{y}}$, $\hat{\sf y} \rightarrow \hat{\sf{x}}$ and $\hat{\sf z} \rightarrow -\hat{\sf{z}}$ so the interaction along the $\sf{y}$ bond emerging out of site 1 in matrix notation is \begin{equation}\mathcal{J}^{\sf y}=\left(\begin{array}{ccc} J' & 0 & -\Gamma' \\ 0 & J'+K' & 0 \\ -\Gamma' & 0 & J' \end{array}\right),\label{eq:Jy_xyz}\end{equation} with \begin{equation}\mathcal{J}^{\sf y}_{abc}=\left(\begin{array}{ccc} J'+\frac{K'}{2} & -\frac{\Gamma'}{\sqrt{2}} & -\frac{K'}{2} \\ -\frac{\Gamma'}{\sqrt{2}} & J' & -\frac{\Gamma'}{\sqrt{2}} \\ -\frac{K'}{2} & -\frac{\Gamma'}{\sqrt{2}} & J'+\frac{K'}{2} \end{array}\right).\label{eq:Jy}\end{equation} The exchange matrix in (\ref{eq:Jy_xyz}) is not simply obtained from the exchange matrix in (\ref{eq:Jx_xyz}) by cyclic permutation of the $\sf{xyz}$ labels, but the off-diagonal term changes sign via the symmetry operation that relates the two bonds, as noted in Ref.~\cite{lee2015hyperhoneycomb}. The off-diagonal exchange also changes sign between bonds of the same colour between chains running along the two distinct $\bmm{a}\pm\bmm{b}$ directions, for example in Fig.~\ref{fig:diffraction}g between the (red) 1-3 and 2-4 $\sf{x}$-bonds related by a 2$_y$ axis passing through the middle of the connecting 1-2 bond. Bonds of the same colour on the same or parallel zigzag chains are identical as they are related by $F$-centring translations.

{\bf Energetic selection of spin orientations on $\sf{x}$- and $\sf{y}$-bonds.} By analogy with the energetic selection discussed in the case of the $\sf{z}$-bond in (\ref{eq:Jz_xyz}), for the $\sf{x}$-bond with exchange (\ref{eq:Jx_xyz}) in the case $K'=0$, $J'>0$ and $\Gamma'>0$, the directions in order of increasing energy are $\hat{\sf{y}}+\hat{\sf{z}}$, $\hat{\sf{x}}$, $\hat{\sf{y}}-\hat{\sf{z}}$. Similarly, for the $\sf{y}$-bond in eq.~(\ref{eq:Jy_xyz}), the directions in order of increasing energy as $\hat{\sf{z}}-\hat{\sf{x}}$, $\hat{\sf{y}}$, $\hat{\sf{z}}+\hat{\sf{x}}$. Because the above energetically most favourable directions are different between $\sf{x}$- and $\sf{y}$-bonds that share a site, a compromise must be reached. The mean-field calculation developed in the following section shows that the compromise energetically-preferred direction is $-\hat{\sf{x}}+\hat{\sf{y}}+\sqrt{2}\hat{\sf{z}}$, parallel to $\bmm{\hat{a}}-\bmm{\hat{b}}$, i.e. at $45^{\circ}$ to the $\bmm{a}$ and $\bmm{b}$ axes. This analysis applies to the zigzag chain containing sites 1 and 3 in Fig.~\ref{fig:diffraction}g, which is representative of zigzag chains oriented along the $\bmm{a}-\bmm{b}$ direction. For the zigzag chains running along the $\bmm{a}+\bmm{b}$ diagonal the compromise energetically-preferred direction is rotated 90$^{\circ}$ to be along $\bmm{\hat{a}}+\bmm{\hat{b}}$.

\section{Mean-Field Model of the Magnetic Structure}
\label{sec:ground_state}
\vspace{1.5mm}
{\bf Mean-field model.} Starting with $J,J'>0$ (antiferromagnetic) the magnetic structure has ordered spins collinear and antiparallel on all bonds, i.e. an $A$ basis vector with all spin orientations degenerate. By adding an exchange $\Gamma<0$ on the ${\sf z}$-bonds, the $a$-axis becomes energetically favoured as per eq.~(\ref{eq:Jz}), i.e. the magnetic structure becomes $A_x$. Inspection of eqs.~(\ref{eq:Jx}) and (\ref{eq:Jy}) shows that a $\Gamma'$ term couples $S_x$ and $S_y$ spin components, so a finite $\Gamma'$ tilts the spins away from the $a$-axis towards $b$. The energetically favoured basis vector of the $S_y$ components is $C$ as it has antiferromagnetic alignment on both the ${\sf x}$- and ${\sf y}$-bonds, so both of those bonds gain energy from the $\Gamma'$ exchange. The ground state becomes ($A_x,\pm C_y$) with the mean-field energy (per site)
\begin{eqnarray}
\frac{E(\phi)}{S^2/2}&=&(-J-2J'+\Gamma)\cos^2\phi+(J-2J')\sin^2\phi \nonumber \\
& & \pm 2\sqrt{2}\Gamma'\sin \phi \cos \phi \nonumber
\end{eqnarray}
with $\phi>0$ the tilt angle of the spins away from the $a$-axis and the upper (lower) sign in the ground state basis vectors combination and the energy expression chosen for $\Gamma'<0$ ($\Gamma'>0$). Minimising $E(\phi)$ gives the equilibrium tilt angle $\phi$ in terms of exchanges as
\begin{equation}
\tan 2 \phi=\frac{2\sqrt{2}|\Gamma'|}{2J-\Gamma},\label{eq:phi}
\end{equation}
and the minimum energy
\begin{equation}\frac{E_0}{S^2/2}=-2J'+\frac{\Gamma}{2}-\sqrt{\left(J-\frac{\Gamma}{2}\right)^2 +2\Gamma'^2}.\label{eq:E0}\end{equation}
Setting $\Gamma'=0$ gives $\phi=0$ and $E_0=(-J-2J'+\Gamma)S^2/2$, i.e. recovers the pure $A_x$ magnetic structure selected by $J,J'>0$ and $\Gamma<0$. On the other hand, setting $\Gamma=0$ and $\Gamma'>0$ makes the experimentally determined magnetic structure ($A_x,-C_y$) with $M_y/M_x=\tan \phi$, degenerate with another structure $(C_x,-A_y)$ with $M_x/M_y=\tan \phi$, with $\tan 2 \phi=\sqrt{2}\Gamma'/J$ in both cases.

{\bf Non-collinear order from frustration of $\Gamma$ and $\Gamma'$ exchanges.} The physical interpretation of $\Gamma<0$ is that antiferromagnetically-aligned spins on the ${\sf z}$-bonds prefer to be in the $ab$ plane, closest to the $a$-axis. $\Gamma'>0$ on the $\sf{x}$- and $\sf{y}$-bonds means that antiferromagnetically aligned spins prefer to be also in the $ab$ plane, but closest to the $\bmm{\hat{a}}\pm\bmm{\hat{b}}$ directions, for the zigzag chains running along the $\bmm{a}\pm\bmm{b}$, respectively, as we show below. Those energetically most favourable spin directions are mutually incompatible, with the consequence that a compromise is reached. Starting with $\Gamma<0$ and $\Gamma'=0$ the magnetic structure is $A_x$ with spins along $a$ as preferred by the $\sf{z}$-bonds. Switching on $\Gamma'>0$ rotates the spins at the ends of each $\sf{z}$-bond around the bond axis, in opposite senses for the two ends such as to bring the spins closer to the directions preferred by the two different types of zigzag chains at the bond ends. The above spin rotations keep the spins on the zigzag chains antiparallel, which maximises the energy gain from the $J'$ exchange. Through those spin rotations both $\sf{x}$- and $\sf{y}$-bonds gain energy via $\Gamma'$, more than the energy lost on the $\sf{z}$ bonds due to spins at the two ends rotating away from the optimal antiparallel alignment along $a$ favoured by $\Gamma$. Focusing on the 1-2 $\sf{z}$-bond in Fig.~\ref{fig:diffraction}g, upon switching on $\Gamma'>0$ the spin on site 1 rotates away from $\bmm{a}$ by an angle $\phi$ towards $\bmm{\hat{a}}-\bmm{\hat{b}}$, whereas spin 2 rotates by the same angle away from $-\bmm{a}$ towards the $-(\bmm{a}+\bmm{b})$ direction. According to eq.~(\ref{eq:phi}), the tilt angle $\phi$ increases monotonically from 0 upon increasing $\Gamma'$ and eventually reaches the maximum value $\pi/4$ in the limit of large $\Gamma'\gg J,-\Gamma$. In this limit the zigzag chains are effectively decoupled, and each chain has collinear antiferromagnetic order along its own preferred direction, obtained from $\phi=\pi/4$ as $\bmm{\hat{a}}\pm\bmm{\hat{b}}$ for chains running along the $\bmm{a}\pm\bmm{b}$ directions.

{\bf Relation to the phase diagram of the $JK\Gamma$ model.} We note that the experimentally-determined magnetic structure is not contained in the magnetic phase diagram of the $JK\Gamma$ model under the simplifying assumption that all bonds are symmetry-equivalent, i.e. $J'=J$, $K'=K$ and $\Gamma'=\Gamma$. In particular, in this model $J>0$ and $\Gamma<0$ select the in-phase ($A_x,C_y$) ground state (labelled $AF_a$ in Ref.~\cite{lee2015hyperhoneycomb}), the out-of-phase ground state ($A_x,-C_y$) can only be obtained in a form where the dominant basis vector is $C_y$, not $A_x$ (structure labelled $SS_b$ in Ref.~\cite{lee2015hyperhoneycomb}) in a region in parameter space where $-K \geq J>0$ and $\Gamma>0$; the powder-averaged spinwave spectrum for a representative set of exchange values from that region of parameter space is shown in Supplementary Fig.~\ref{fig:JKG}, which differs qualitatively from the experimentally observed spectrum in Fig.~\ref{fig:spinwave}a.

\begin{figure}
\includegraphics[width=8.5cm]{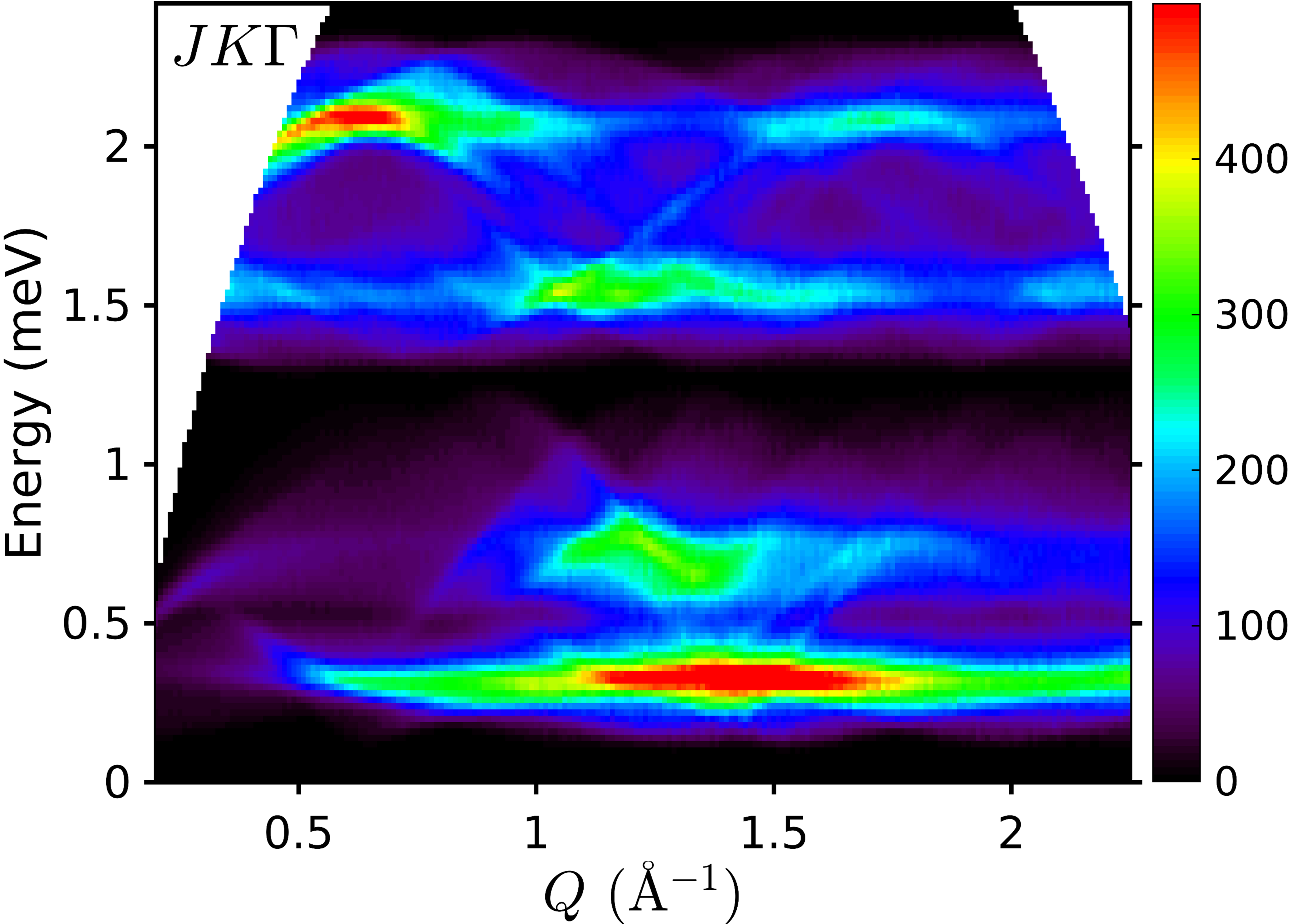}
\caption{\label{fig:JKG} Powder-averaged spinwave spectrum for the $JK\Gamma$ Hamiltonian for which the spectrum along a high-symmetry path in reciprocal space is plotted in Supplementary Fig.~\ref{fig:sw_spectrum}e. The calculation includes magnetic form factor, neutron polarization and convolution with the estimated experimental energy resolution and should be compared to the experimental data in Fig.~\ref{fig:spinwave}a. The colour bar indicates the calculated scattering intensity in arbitrary units on a linear scale.}
\end{figure}

{\bf Comparison to a Heisenberg model with Dzyaloshinskii Moriya (DM) interactions.} For completeness we note that a non-collinear magnetic structure could in principle also be stabilized by an antiferromagnetic Heisenberg exchange $J$ on all nearest-neighbour bonds and a symmetry-allowed DM interaction on the $\sf{z}$-bonds, $\bmm{D}\! \cdot\! \left( \bmm{S}_i \mathbf{\times} \bmm{S}_j \right)$ where $\bmm{D} \parallel \bmm{c}$ and $i,j$ index the lower/upper sites on all $\sf{z}$-bonds, which are 2-fold rotation axes of the crystal structure. However, to quantitatively reproduce the experimentally deduced non-collinearity would require $D/J=\tan{2\phi}\simeq1.5$ (using the experimentally determined $\phi=\arctan(M_y/M_x)$ and assuming an isotropic $g$-tensor in the $ab$ plane). Such a large value of $D/J$ is unphysical as $D$ is typically a subleading exchange, arising from perturbative inclusion of the spin-orbit coupling. Furthermore, this $JD$ Hamiltonian has rotational $U(1)$ symmetry around the $z$ axis, so the moments in the ground state could be continuously rotated together in the $ab$ plane around the $z$-axis at no energy cost via a gapless Goldstone mode, contrary to the observation of a clearly gapped magnetic spectrum in Fig.~\ref{fig:spinwave}a. For those reasons we conclude that the $J\Gamma\Gamma'$ model discussed previously is a more likely minimal model consistent with all experimentally observed key features of the magnetic order and dynamics.

\section{Spinwave spectrum}
\label{sec:spinwave}
\vspace{1.5mm}
{\bf Primitive cell.} Calculations of the spin-wave spectrum for model Hamiltonians were performed using SpinW \cite{spinW} in the primitive magnetic cell, which coincides with the primitive structural cell, with basis vectors related to the conventional orthorhombic cell vectors by
$$\left(\begin{array}{c} \bmm{a}_p\\ \bmm{b}_p \\ \bmm{c}_p\end{array}\right)
=\frac{1}{2}\left(\begin{array}{ccc} 0 & 1 & 1 \\ 1 & 0 & 1 \\ 1 & 1 & 0 \end{array}\right) \left(\begin{array}{c} \bmm{a}\\ \bmm{b} \\ \bmm{c}\end{array}\right).$$ There are four magnetic sublattices and six bonds per primitive cell, two each of type $\sf{x}$, $\sf{y}$ and $\sf{z}$, as illustrated in Supplementary Fig.~\ref{fig:primitive}a. For each of the two $\sf{z}$-bonds the exchange matrix is $\mathcal{P}^T\mathcal{J}^{\sf z}_{abc}\mathcal{P}$, where $\mathcal{P}$ is the transformation matrix between the orthorhombic unit cell vectors and the $\mathscr{XYZ}$ SpinW Cartesian spin axes associated with the primitive cell,
$$\left(\begin{array}{c} \bmm{\hat{a}}\\ \bmm{\hat{b}} \\ \bmm{\hat{c}}\end{array}\right)
=\mathcal{P} \left(\begin{array}{c} \bmm{\hat{\mathscr{X}}}\\ \bmm{\hat{\mathscr{Y}}} \\ \bmm{\hat{\mathscr{Z}}}\end{array}\right),$$
where $\bmm{\hat{\mathscr{X}}}=\bmm{a}_p/|\bmm{a}_p|$, $\bmm{\hat{\mathscr{Z}}}=\bmm{a}_p \times \bmm{b}_p/|\bmm{a}_p \times \bmm{b}_p|$ and $\bmm{\hat{\mathscr{Y}}}=\bmm{\hat{\mathscr{Z}}} \times \bmm{\hat{\mathscr{X}}}$. The exchange matrix for the representative $\sf{x}$-bond (1-3 on the $\bmm{a}-\bmm{b}$ chain in Fig.~\ref{fig:primitive}a is similarly obtained as $\mathcal{P}^T\mathcal{J}^{\sf x}_{abc}\mathcal{P}$ and the rest of the (four) exchange bonds in the primitive cell are obtained via symmetry operations of the crystal structure (2$_y$ and 2$_z$ rotations passing through the middle of the 3-4 $\sf{z}$-bond).

\begin{figure}
\includegraphics[width=8 cm]{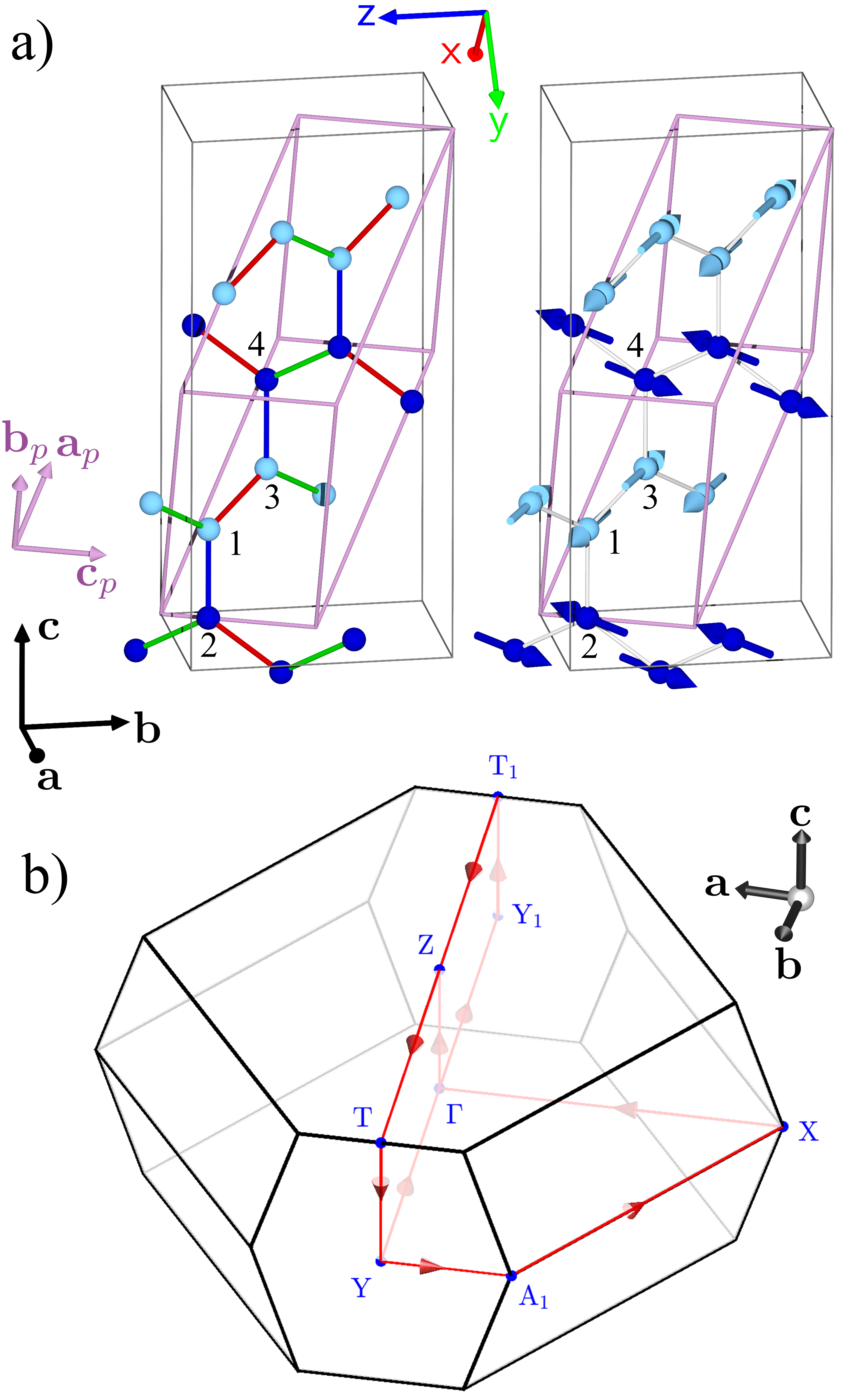}
\caption{\label{fig:primitive}\textbf{a,} Primitive (magenta) vs. conventional cell (thin black solid outline). Labels 1-4 show sites equivalent to those listed in Supplementary Table~\ref{tab:BV} up to $F$-centring translations. \textbf{b,} Brillouin zone with $\Gamma$X, $\Gamma$Y, $\Gamma$Z along the $-\bmm{a}$, $\bmm{b}$ and $\bmm{c}$ axes, respectively. High-symmetry points connected by the thick red arrowed lines show the path along which the spinwave spectrum is plotted in Supplementary Fig.~\ref{fig:sw_spectrum}, which contains also the ($hkl$) indices of the labelled points.}
\end{figure}

{\bf Brillouin zone.} The Brillouin zone corresponding to the primitive cell is illustrated in Supplementary Fig.~\ref{fig:primitive}b and belongs to the $F$-centred orthorhombic unit cells with $1/a^2>1/b^2+1/c^2$ \cite{Bradley}. It has top and bottom distorted hexagonal faces with midpoints at $\pm(001)$, normal side distorted hexagonal faces with midpoints at $\pm(010)$ and additional eight slanted rectangular side faces with midpoints at $\left(\frac{1}{2},\frac{1}{2},\frac{1}{2}\right)$ and symmetry-equivalent positions obtained by $mmm$ structural point group operations.

{\bf Key features of the spinwave spectrum.} Four magnon modes are expected at a general wavevector, equal to the number of magnetic sublattices, but additional degeneracies occur in special cases. Starting with the simplest case $J=J'>0$ and $\Gamma=\Gamma'=0$, the ground state is collinear N\'{e}el ordered in an $A$ basis vector with polarization selected via spontaneous symmetry-breaking with a linearly-dispersing gapless Goldstone mode emerging out of each $\Gamma$-point zone centre, as shown in the Supplementary Fig.~\ref{fig:sw_spectrum}a. At a general wavevector there are two doubly-degenerate magnon branches, with degeneracy protected by the rotational symmetry of the spin Hamiltonian.

Upon switching on an off-diagonal exchange $\Gamma>0$ on the $\sf{z}$-bonds the continuous rotational symmetry is broken and the ground state basis vector $A_x$ is energetically selected, i.e. the magnetic structure remains collinear N\'{e}el ordered, but ordering breaks now a discrete Ising symmetry and as a consequence the spectrum has a gap above the magnetic Bragg peaks, scaling to leading order as $\sqrt{-\Gamma J}$. The breaking of rotational symmetry removes the two-fold degeneracy of the magnon branches with four non-degenerate modes at a general wavevector, as illustrated in the Supplementary Fig.~\ref{fig:sw_spectrum}b. The spectrum is identical between time-reversed domains, i.e. $\pm A_x$, and mirrored in $h$, $k$ and $l$, as the spin Hamiltonian is invariant under reversal of any of $x$, $y$ or $z$ axes.

\begin{figure*}
\includegraphics[width=15.7cm]{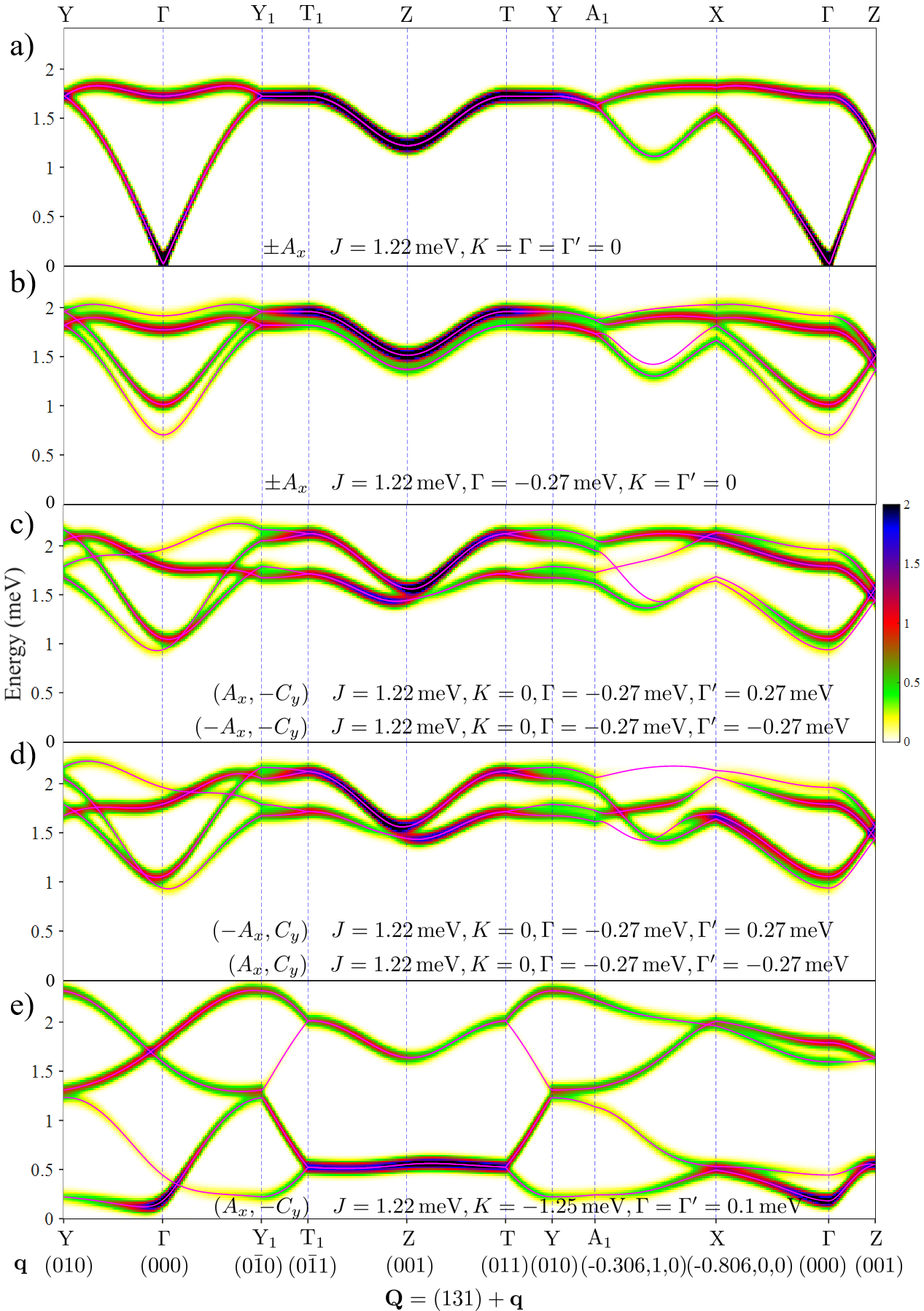}
\caption{\label{fig:sw_spectrum} Spinwave spectrum for various spin Hamiltonians discussed in the text, with magnetic domain and exchange parameters listed in each panel with $K'=K$ and $J'=J$. The wavevector follows the high-symmetry path in the Brillouin zone shown by the thick arrowed lines in Supplementary Fig.~\ref{fig:primitive}b, offset to the (131) zone centre chosen since almost all modes carry finite intensity. Horizontal top and bottom labels indicate special high-symmetry points in the Brillouin zone, also labelled in Fig.~\ref{fig:primitive}b, and the indices show the reduced wavevector $\bmm{q}$ in the Brillouin zone. Solid lines are the magnon dispersion relations and colour is the dynamical correlation $S^{zz}(\bmm{Q},\omega)$ after convolution with a Gaussian in energy of FWHM 0.1~meV (no magnetic form factor contribution). The powder-averaged spectrum for the top model in panels c and e is shown in Fig.~\ref{fig:spinwave}b and Supplementary Fig.~\ref{fig:JKG}, respectively. The colour bar indicates the intensity in arbitrary units on a linear scale.}
\end{figure*}

Upon additionally switching on a finite $\Gamma'>0$ the structure becomes non-collinear by mixing into the ground state a $-C_y$ basis vector. The spectrum was already gaped with modes split, the addition of $\Gamma'$ further increases the gap and also modifies the magnon energies such that the spectrum is no longer mirrored in $k$ and time-reversed domains no longer have identical dispersion relations, as discussed in the following paragraph. A two-fold symmetry-protected degeneracy of the magnon bands is still preserved for the Brillouin zone boundary path T$_1$ZT as shown in Fig.~\ref{fig:sw_spectrum}c.

{\bf Symmetry of the spinwave spectrum in reciprocal space.} As discussed in \ref{sec:structure_factors} the magnetic point group symmetry is $m'mm$ ($m'$ normal to $\bmm{b}$). Under operators of this point group a general wavevector $\bmm{Q}=(hkl)$ is mapped into itself, or into ($\bar{h}k\bar{l}$), ($hk\bar{l})$ and ($\bar{h}kl)$. Therefore dispersion relations are mirrored in $h$ and $l$, but not necessarily in $k$. Indeed this is shown in Supplementary Fig.~\ref{fig:sw_spectrum}c, note that the left-right symmetry of the dispersion relations along the paths Y$\Gamma$Y$_1$ and T$_1$ZT are broken (the $k$-component of the reduced wavevector $\bmm{q}$ switches sign in the middle of each of those two paths), i.e. left- and right-moving spin waves propagating along the (010) direction are non-reciprocal, with distinct dispersion relations. However, symmetry operations that are broken at the magnetic phase transition map one magnetic domain onto its time-reversed counterpart. The same operators map $k$ into $\bar{k}$, therefore the dispersions of the time-reversed domain at wavevector ($h\bar{k}l$) are the same as the dispersions of the original domain at ($hkl$), as apparent by comparing Supplementary Fig.~\ref{fig:sw_spectrum}c and d. The above mapping between the spectra of time-reversed domains has the consequence that for calculating the spherically averaged spectrum over wavevector orientations, relevant for comparison with the powder inelastic neutron scattering spectrum, it is sufficient to consider only one magnetic domain, as its time-reversed pair has an identical spherically-averaged spectrum. The non-reciprocal nature of the spinwave spectrum along (010) has the consequence that in a macroscopic single-crystal sample where both time-reversed magnetic domains co-exist, one would expect eight distinct spin wave modes (four for each of the two domains) at a general position in reciprocal space where the reduced wavevector has a finite $k$-component.

{\bf Transformation of the spinwave spectrum under sign reversal of the $\Gamma'$ term.} An important property of the Hamiltonian is that reversing the sign of $\Gamma'$ is identical to reversing the $y$-axis, as can be seen by inspecting eqs.~(\ref{eq:Jz}), (\ref{eq:Jx}) and (\ref{eq:Jy}). This has the consequence that reversing the sign of $\Gamma'$ reverses the sign of the $S_y$ components in the magnetic ground state, but the magnitude of the tilt angle $\phi$ and the ground state energy remain unchanged, see eqs.~(\ref{eq:phi}) and (\ref{eq:E0}). Consider for concreteness the spectrum of the $\Gamma'>0$ model for the ($A_x,-C_y$) magnetic domain at wavevector ($hkl$). Reversal of the $y$-axis maps the Hamiltonian into the case of sign reversed $\Gamma'$, the magnetic domain into ($A_x,C_y)$ and the wavevector into ($h\bar{k}l$). Applying now time reversal leaves the Hamiltonian invariant, maps the magnetic domain into its time-reversed counterpart ($-A_x,-C_y)$ and the wavevector into ($\bar{h}k\bar{l}$), which is equivalent to ($hkl$) via the magnetic point group operations. Therefore, the dispersions of the $\Gamma'>0$ model and magnetic domain ($A_x,-C_y$), and the sign reversed $\Gamma'$ and magnetic domain ($-A_x,-C_y$) are identical. The components of the dynamical correlations that contain one polarization along the $y$-axis, i.e. $S^{xy}(\bmm{Q},\omega)$, $S^{yx}$, $S^{yz}$, $S^{zy}$ change sign, whereas all other components $S^{xx}$, $S^{yy}$, $S^{zz}$, $S^{xz}$ and $S^{zx}$ are unchanged, therefore the spherically-averaged spectrum is quite similar, but not identical between the two cases.

\end{document}